\documentclass[twocolumn]{elsarticle}
\usepackage{lineno,hyperref}
\modulolinenumbers[5]
\usepackage{graphicx,amssymb}
\usepackage{abstract}
\journal{Journal of High Energy Astrophysics}

\usepackage[misc]{ifsym}
\newcommand\blfootnote[1]{%
	\begingroup
	\renewcommand\thefootnote{}\footnote{#1}%
	\addtocounter{footnote}{-1}%
	\endgroup
}

\usepackage[hang]{footmisc}

\begin{document}

\title{Studying magnetic fields of ultraluminous X-ray pulsars using different accretion torques}


\author[1,2]{X. Chen}
\author[1,2]{W. Wang\corref{mycorrespondingauthor}}
\cortext[mycorrespondingauthor]{Corresponding author}
\author[3]{H. Tong}

\address[1]{School of Physics and Technology, Wuhan University, Wuhan 430072, China}
\address[2]{WHU-NAOC Joint Center for Astronomy, Wuhan University, Wuhan 430072, China}
\address[3]{School of Physics and Materials Science, Guangzhou University, Guangzhou 510006, China}

\twocolumn[
\maketitle 

\begin{onecolabstract}

	The magnetic field of ultraluminous X-ray (ULX) pulsars is the key parameter to understand the nature and accretion physics. However, the typical magnetic field values in these ULX pulsars are still under debate. We used six different torque models to study the magnetic fields of ULX pulsars, to see how derived magnetic fields change with different models, and to determine which models are more suitable for ULX pulsars. We took the currently available period, period derivative, and flux data of 7 confirmed ULX pulsars, M82 X-2, ULX NGC 7793 P13, ULX NGC 5907, NGC 300 ULX1, NGC 1313 X-2, M51 ULX-7, Swift J0243.6+6124, plus one potential ULX pulsars, SMC X-3. The magnetic fields of these ULX pulsars were constrained from two physical conditions: the spin-up process and near equilibrium. We checked possible dependence of the magnetic field estimations on the different torque models. The calculations suggest the accretion torque models by Ghosh \& Lamb \cite{Ghosh1979}, Wang \cite{Wang1995}, Klu$\rm \acute{z}$niak \& Rappaport \cite{Kluzniak2007} and Campbell \cite{Campbell2012} are more likely to support magnetar models for ULX pulsars, while Lovelace, Romanova \& Bisnovatyi-Kogan \cite{Lovelace1995}'s model generally predicts the magnetic field in normal neutron stars. Implications of our results combined with other independent methods are also discussed, which will help us to understand the nature and rotational behavior of these ULX pulsars.
	\newline
	
\end{onecolabstract}

]
\author[1,2]{X. Chen}
\author[1,2]{W. Wang}
{
	\blfootnote{
		\hspace{-0.15in} *Corresponding author
		\newline  Email address: wangwei2017@whu.edu.cn (W. Wang)
		\newline}
}

\author[3]{H. Tong}
\section{Introduction}\label{section1}
Ultraluminous X-ray sources are off-nuclear objects with extremely high luminosities in near galaxies. For a long time, they were thought to be intermediate mass black holes \cite{Colbert1999,Liu2008,King2001,Gladstone2009}, but the recent discovery of pulsation gave the evidence that at least some of them are neutron stars. Up to now, 7 sources are believed to be ULX pulsars: M82 X-2 \cite{Bachetti2014}, NGC 7793 P13 \cite{Furst2016,Israel2017b}, NGC 5907 ULX1 \cite{Israel2017a}, NGC 300 ULX1 \cite{Carpano2018}, NGC 1313 X-2 \cite{Sathyaprakash2019}, M51 ULX7 \cite{RodriguezCastillo2020}, and a source within the Milky Way, Swift J0243.6+6124. XMMU J031747.5-663010 in NGC 1313 \cite{Trudolyubov2008}, SMC X-3 \cite{Tsygankov2017,Weng2017} and M51 ULX8 \cite{Brightman2018} may also be ULX pulsars. Recently, Song et al. \cite{Song2020} identified 25 sources with their flux changes of more than one order of magnitude from 296 ULXs, and claimed these are possible ULX pulsar candidates.

The luminosities of these ULX pulsars are much higher than those normal X-ray pulsars in the Milk Way, but the physical origin of such extreme luminosities is still not clear. Based on their similarly sinusoidal pulse profiles \cite{Bachetti2014,Furst2016,Israel2017a,Israel2017b,Carpano2018}, they are not likely to be strongly beamed, even though anisotropy must exist \cite{Basko1976}. However, their magnetic fields remain debatable. Some authors have found evidences of strong magnetic fields as large as $10^{15} \,\rm G$ \cite{Duncan1992,Kaspi2017,Brightman2018,Middleton2019}, but some found moderate fields of around $10^{12} \,\rm G$ \cite{Walton2018}, or even weaker \cite{Kluzniak2015}.

We assume that these ULX pulsars are strongly magnetized accretion neutron stars \cite{Tong2015,Tong2019}, and apply the existing accretion torque models to these sources to analyze the accretion process and the ultraluminous phenomenon generation conditions. The models used here are thin disk accretions, including the basic Ghosh and Lamb \cite{Ghosh1979} model and the revised versions with different physical conditions. They have been well studied and applied to the evolution of neutron stars in binary systems and X-ray pulsars \cite{Shakura2012,Ho2014,Wang2020}, but have not been discussed in details for ULX pulsars. We infer magnetic fields from different model torques, and substitute the observation data (e.g., spin period and its derivative, distance, and flux) to analyze how magnetic fields given by different torque models vary. Even though the observational data is not sufficient enough, we use the current data to discuss which torque form is more suitable for these pulsating ULXs.

We organize this paper in the following form. In section 2 we introduce the accretion torque models and the calculation process. Section 3 gives the results of magnetic field by applying different accretion torque models based on the existing ULX pulsar data, and the discussions and conclusions are presented in Section 4.

\section{The accretion torque models}\label{sec:2}
Six torque models are utilized to calculate the magnetic field, including Ghosh \& Lamb \cite{Ghosh1979}'s classical model (hereafter GL), Wang \cite{Wang1995}'s model (containing 2 derived magnetic field forms, hereafter WangAB and WangC), Lovelace, Romanova \& Bisnovatyi-Kogan \cite{Lovelace1995}'s model (hereafter LRB), Klu$\rm \acute{z}$niak \& Rappaport \cite{Kluzniak2007}'s model(hereafter KR), and Campbell \cite{Campbell2012}'s model (hereafter Camp). These models are thin-disk accretion models, which are basically modified versions based on the GL model.

The inner radius of the disk, $R_{\rm in}$, is determined by the balance of magnetic and material pressure \cite{Ghosh1979}. Based on the conservation of angular momentum, we write
\begin{equation}\label{equation:Rin}
	\frac{d}{dR}(\dot{M}R^{2}\Omega)=-R^{2}B^+_{\rm \phi} B_{\rm z},
\end{equation}
where $\dot{M}$ is the mass accretion rate, $\Omega$ is the angular velocity, $B^+_{\rm \phi}$ is the toroidal magnetic field, and $B_{\rm z}$ is the poloidal field. Integrating the above equation over a very narrow region from $R_{\rm in}-\Delta R$ to $R_{\rm in}$, we can obtain the inner radius of the disk. Note that in the calculation, $\Omega$ at $R_{\rm in}-\Delta R$ and $R_{\rm in}$ are assumed to be the angular velocity of the neutron star $\Omega_{\rm *}$ and the Keplerian angular velocity at the inner radius $\Omega_{\rm K}(R_{\rm in})$, respectively. $B_{\rm z}$ is assumed as $-\mu/R^{3}$, where $\mu$ is the magnetic dipole moment, and the mass accretion rate is steady through the innermost disk and onto the center star, thus
\begin{equation}\label{equation:Mdot}
	\dot{M}_{\rm in}=\dot{M}_{\rm *}=\frac{4 \pi b R_{\rm *} d^{2} F_{\rm x}}{G M_{\rm *}},
\end{equation}
where $\dot{M}_{\rm *}$ is the mass accretion rate onto the star, b presents the beaming fraction, $R_{\rm *}$ is the neutron star radius, $M_{\rm *}$ is the neutron star mass, d is the source distance, and $F_{\rm x}$ presents the X-ray flux.

The fastness parameter $\omega_{\rm *}$ is defined as
\begin{equation}\label{equation:w*}
	\omega_{\rm *} = \frac{\Omega_{\rm *}}{\Omega_{\rm K}(R_{\rm in})} = (\frac{R_{\rm in}}{R_{\rm co}})^{3/2},
\end{equation}
where the corotation radius $R_{\rm co}$ is related to the spin period $P$, i.e., $R_{\rm co} = (\frac{G M_{\rm *}}{\Omega_{\rm *}^2})^{1/3} = (\frac{G M_{\rm *}}{(\frac{2\pi}{P})^2})^{1/3}$. Substituting the derived $R_{\rm in}$ into Equation \ref{equation:w*}, the magnetic dipole field can be expressed. Since the $B_{\rm \phi} - B_{\rm z}$ relation in each model are different, the inner radii and the magnetic fields also have different expressions, and are listed in details in \ref{GL_model} - \ref{Camp_model} below.

The accretion torque is believed to affect the spin period, and has the relation of $N=-\dot{P} 2 \pi I / P^{2}$ \cite{Ghosh1977}, where $I$ is the moment of inertia. Combining with the following form of accretion torque, $N=nN_{\rm material}$, we can infer the formulas of $B$ and $\omega_*$ with respect to period, period derivative, X-ray flux, distance, beaming fraction, radius of star, mass of star and moment of inertia backwards from the torque expressions, where $n$ is dimensionless torque. Here the material torque $N_{\rm material}=\dot{M}_{\rm *} \sqrt{GM_{\rm *}R_{\rm in}}$ is utilized in the model calculation except Camp model (see below \ref{Camp_model} for the details). If the system is away from spin equilibrium, then the dimensionless torque is approaching a constant of order unity, and the limitation $\omega_{\rm *} \rightarrow 0$ is adopted to calculate $n$. While the critical fastness parameter $\omega_{\rm crit}$ is calculated by setting $n=0$ for the equilibrium regime. The six torque models and their formulas deriving the magnetic fields are listed as follows in details:

\begin{enumerate}
	\item
	{\bf GL model}. GL model (or its simplified version) is the most commonly used model. It includes an outer transition zone where the magnetic lines cross through the accretion disk and the angular velocity is Keplerian. Then a boundary layer is in the internal bordering the outer zone at $R_{\rm 0}$, where the angular velocity starts to transform to stellar type and converted rapidly at the corotation radius, $R_{\rm co}$, which is also the inner bound of the boundary layer. Materials here flows along the magnetic field lines from the disk to the accretion star.
	
	Since $R_{\rm in} \approx 0.52\mu^{4/7}(2GM_{\rm *})^{-1/7}\dot{M}^{-2/7}_{\rm *}$ is given in Ghosh \& Lamb \cite{Ghosh1979}'s paper, combined with Equation \ref{equation:w*}, the dipole magnetic field can be obtained. If the torque system is near equilibrium, then $\omega_{\rm *}$ can be set as a constant critical value, thus can be used for the spin equilibrium magnetic field calculation, which is given by
	\begin{equation}\label{equation:B_GL}
		B_{\rm eqm}=\frac{2^{1/12} \omega_{\rm *}^{7/6} \dot{M}^{1/2}_{\rm *} (GM_{\rm *})^{5/6} P^{7/6}}{0.52^{7/4} \pi^{7/6} R^3_{\rm *}}.
	\end{equation}
	In GL model, $\omega_{\rm *} = \omega_{\rm crit} = 0.35$ when $n=0$.
	
	Additionally, when considering the spin up regime, the corresponding fastness parameter and the magnetic field can be obtained substituting Equation \ref{equation:Mdot}, Equation \ref{equation:w*} and $R_{\rm in}$ into
	\begin{equation}\label{equation:torque}
		N=-\dot{P} 2 \pi I / P^{2}=nN_{\rm material},
	\end{equation}
	which can lead to the following solution:
	\begin{equation}\label{equation:wGL}
		\omega_{\rm *}=(\frac{4 \pi^2}{G M_{\rm *}})^2 (\frac{I |\dot{P}|}{n \dot{M}_{\rm *} P^{7/3}})^{3}
	\end{equation}
	and
	\begin{equation}
		B_{\rm spinup}=\frac{\omega_{\rm *}}{R^{3}_{\rm *}} \sqrt{\frac{2 G M_{\rm *} I |\dot{P}|}{\pi n}} C,
	\end{equation}
	where $C=(0.52*2^{-1/7})^{-7/4}$ is a constant. If $\omega_{\rm *} \rightarrow 0$, $n \approx 1.39$ is adopted in the equations.
	\label{GL_model}
	
	\item {\bf WangAB model}. Wang \cite{Wang1995} argued that the toroidal magnetic field $B_{\rm \phi}$ is higher than that in GL model. He has recalculated it with three different methods (the first two methods named as Wang AB model, while the third method as WangC model). For the Wang AB model, the dissipation timescale for $B_{\rm \phi}(R)$, i.e. $\tau_{\rm \phi}$, is determined by turbulent diffusion in the disk and by reconnection outside the disk, respectively, but have the same expression when $R$ is approaching $R_{\rm in}$, thus the magnetic field expressions are the same. Following the description above, $R_{\rm in}$ can be derived from Equation \ref{equation:Rin} as
	\begin{equation}
		R_{\rm in} = \frac{B^{4/7}R^{12/7}_{\rm *}\delta^{2/7}}{2^{4/7}\dot{M}_{\rm *}^{2/7}(GM_{\rm *})^{1/7}},
	\end{equation}
	where $\delta=\Delta R/R_{\rm in}$. Then the same derivation method as GL model can be used to get the equilibrium state expression as
	\begin{equation}
		B_{\rm eqm}=\frac{\omega_{\rm *}^{7/6} \dot{M}^{1/2}_{\rm *} (GM_{\rm *})^{5/6} P^{7/6}}{2^{1/6} \pi^{7/6} \delta^{1/2} R^3_{\rm *}}.
	\end{equation}
	
	The two methods in WangAB model have different dimensionless torque forms, and thus their critical fastness parameters are different (0.875 and 0.95 respectively). Here we set $\omega_{\rm crit} = 0.95$ just for convenience. Substitute $R_{\rm in}$ into Equation \ref{equation:torque}, $\omega_{\rm *}$ will be the same as Equation \ref{equation:wGL} in GL model, and the magnetic field of the spin-up regime has the form
	\begin{equation}
		B_{\rm spinup}=\frac{\omega_{\rm *}}{R^{3}_{\rm *}} \sqrt{\frac{2 G M_{\rm *} I |\dot{P}|}{\pi n \delta}},
	\end{equation}
	where $n \approx 7/6$ for both methods.
	
	\item {\bf WangC model}. This scenario corresponds to $\tau_{\rm \phi}$ determined by Alfv$\rm \acute{e}$n speed in the disk and contains the thermal pressure $p(R)=p_{\rm 0} (R_{\rm 0}/R)^{51/20}$. The inner radius is
	\begin{equation}
		R_{\rm in} = \frac{B^{3/4}R^{9/4}_{\rm *}C^{1/2}\pi^{1/8}\delta^{1/4}}{2^{1/4}\dot{M}_{\rm *}^{1/2}(GM_{\rm *})^{1/4}},
	\end{equation}
	where $C=\frac{\sqrt{\frac{\gamma}{\xi}}p^{1/4}_{\rm 0}}{\sqrt{3}}$. Here $\gamma \gtrsim 1$ and $\xi<1$ are numerical factors, and we assume $\frac{\gamma}{\xi} \approx 1$ in the calculation. The equilibrium form of magnetic field is
	\begin{equation}
		B_{\rm eqm}=\frac{\omega_{\rm *}^{8/9} \dot{M}^{2/3}_{\rm *} (GM_{\rm *})^{7/9} P^{8/9}}{2^{5/9} \pi^{19/18} C^{2/3} \delta^{1/3} R^3_{\rm *}},
	\end{equation}
	where $\omega_{\rm *} = \omega_{\rm crit} = 0.949$. Again, when considering the spin-up rate, $\omega_{\rm *}$ is the same as Equation \ref{equation:wGL}, and the magnetic field is given by
	\begin{equation}
		B_{\rm spinup}=\frac{{\omega_{\rm *}}^{2/3}}{R^{3}_{\rm *}} (\frac{2 G M_{\rm *} (I |\dot{P}|)^2}{\pi^{1/2} C^2 n^2 \delta P^2})^{1/3},
	\end{equation}
	where $n \approx 1.23$.
	
	\item {\bf LRB model}. Lovelace et al. \cite{Lovelace1995} have built a new model considering the turbulent viscosity of the disc. They argued that when stellar angular velocity is different from disc value, the magnetic field will quickly expand and form open lines. This process will generate outflows. Depending on the relation of the inner radius of the outflow and $r_{\rm co}$, star may spin-up, spin-down, or even fluctuating back and forth in two scenarios. With similar formula derivations, we got the inner radius as
	\begin{equation}
		R_{\rm in} = \frac{B^{1/2}R^{3/2}_{\rm *}C^{1/4}\delta^{1/4}}{2^{1/2}\dot{M}_{\rm *}^{1/4}},
	\end{equation}
	where $C=\frac{3}{2 \alpha D c_{\rm s}}$. Note that $\alpha$ and $D$ are dimensionless quantities, and $\alpha D$ is in from 0.01 to 0.1 \cite{Lovelace1995}, while $c_{\rm s}$ is the Newton sound speed. We assume $\alpha D = 0.1$ in this work. The magnetic field descriptions in the equilibrium scenario and spin-up regime, respectively
	\begin{equation}
		B_{\rm eqm}=\frac{\omega_{\rm *}^{4/3} \dot{M}^{1/2}_{\rm *} (GM_{\rm *})^{2/3} P^{4/3}}{2^{1/3} \pi^{4/3} C^{1/2} \delta^{1/2} R^3_{\rm *}},
	\end{equation}
	and
	\begin{equation}
		B_{\rm spinup}=\frac{{\omega_{\rm *}}^{7/6}}{R^{3}_{\rm *}} (\frac{4 (G M_{\rm *})^2 (I |\dot{P}|)^3 P}{\pi^4 C^3 n^3 \delta^3})^{1/6}.
	\end{equation}
	The $\omega_{\rm *}$ for $B_{\rm spinup}$ is also the same as Equation \ref{equation:wGL}. Lovelace et al. \cite{Lovelace1995} proposed a limit on $B_{\rm \phi} - B_{\rm z}$ relation, and thus gave no dimensionless torque form. Refer to their Equation 18a, we use $n=1$ in our $B_{\rm spinup}$ calculation. While for $B_{\rm eqm}$, we assume their turnover radius $R_{\rm to}=R_{\rm co}$ is the approximate critical condition. Since their $R_{\rm to}$ has a similar role to the Alfv$\acute{e}$n radius $R_{A}$ of GL model, and is smaller than $R_{A}$ by a factor of about 0.24 for $\alpha D=0.1$ \cite{Lovelace1995}, thus $\omega_{\rm crit}$ is estimated to be 0.566.
	
	\item {\bf KR model}. The authors made a modification to the azimuthal magnetic field component and considered its relations with radius. In their model, $B_{\rm \phi}$ is either dissipated by turbulent diffusion or recombination following Wang \cite{Wang1995}. Two different forms of magnetic fields are given when $R<R_{\rm co}$. Since the results are similar, we just adopt the first form, same as the authors mainly discussed in the paper. In this case, the inner radius is
	\begin{equation}
		R_{\rm in} = \frac{B^{2/5}R^{6/5}_{\rm *}P^{1/5}\delta^{1/5}}{2^{3/5}\pi^{1/5}\dot{M}_{\rm *}^{1/5}}.
	\end{equation}
	The derived $\omega_{\rm *}$ still has the same expression as Equation \ref{equation:wGL}, and the equilibrium version and spin-up version of magnetic fields are
	\begin{equation}
		B_{\rm eqm}=\frac{\omega_{\rm *}^{5/3} \dot{M}^{1/2}_{\rm *} (GM_{\rm *})^{5/6} P^{7/6}} {2^{1/6} \pi^{7/6} \delta^{1/2} R^3_{\rm *}},
	\end{equation}
	and
	\begin{equation}
		B_{\rm spinup}=\frac{\omega^{3/2}_{\rm *}}{R^{3}_{\rm *}} \sqrt{\frac{2 G M_{\rm *} I |\dot{P}|}{\pi n \delta}},
	\end{equation}
	respectively, where $\omega_{\rm crit}=20/21$ and n $\approx 10/9$.
	
	\item {\bf Camp model}. The author analyzed the formation of accretion curtain flows when the inner part of the disc is disrupted by a strongly magnetized star. The material will leave disc and become a channelled, curtain flow. The inner radius and the equilibrium formula of magnetic field are shown as
	\begin{equation}
		R_{\rm in} = \frac{B^{2/5} R^{6/5}_{\rm *}P^{1/5} \delta^{1/5} C^{1/5}}{2^{3/5}\pi^{1/5}\dot{M}_{\rm *}^{1/5}}
	\end{equation}
	and
	\begin{equation}
		B_{\rm eqm}=\frac{\omega_{\rm *}^{5/3} \dot{M}^{1/2}_{\rm *} (GM_{\rm *})^{5/6} P^{7/6}}{2^{1/6} \pi^{7/6} \delta^{1/2} C^{1/2} R^3_{\rm *}}
	\end{equation}
	respectively, where $C=\frac{\gamma}{\epsilon_{\rm m}}$, and $\omega_{\rm crit}=(2/3)^{1/2}$. $\gamma$ and $\epsilon_{\rm m}$ are dimensionless quantities, and $C=1$ is assumed during the calculation.
	
	As described by Campbell \cite{Campbell2012}, the accretion torque $N=\omega_{\rm *} \dot{M}_{\rm *} \sqrt{G M_{\rm *} R_{\rm in}}$, which leads to a modification of Equation \ref{equation:torque}, i.e., $\omega_{\rm *} \dot{M}_{\rm *} \sqrt{G M_{\rm *} R_{\rm in}} = -\dot{P} 2 \pi I / P^{2}$. Thus, the derived expression of the fastness parameter is
	\begin{equation}
		\omega_{\rm *}=(\frac{4 \pi^2}{G M_{\rm *}})^{1/2} (\frac{I |\dot{P}|}{\dot{M}_{\rm *} P^{7/3}})^{3/4},
	\end{equation}
	and the magnetic field in the spin-up regime is
	\begin{equation}
		B_{\rm spinup}=\frac{\omega_{\rm *}}{R^{3}_{\rm *}} \sqrt{\frac{2 G M_{\rm *} I |\dot{P}|}{\pi C \delta}}.
	\end{equation}
	\label{Camp_model}
	
\end{enumerate}

\section{Understanding magnetic fields of ULX pulsars}\label{sec:3}
\subsection{Observations of the rotational behaviors for different sources}
There are seven confirmed ULX pulsars at present, including six extragalactic sources and one case in the Galaxy. We also analyze an extra extragalactic sources which are possible ULX pulsars. Table~\ref{tab:input} shows all the important observed parameters of different sources obtained from literatures. We have included the different observations for each source, while the only observed parameters with notable differences are included, and in the calculations, we avoid using $\dot{P}$ computed based on just comparing two spin period or long timescale average measurements (marked with $*$). Since the observation information and parameters are not complete, many other measured data \cite{Bahramian2017,Beri2021,Carpano2018,Vasilopoulos2018b,Vasilopoulos2019,Ray2019,Wilson2018} are not listed.

\begin{table*}
	\scriptsize
	\caption{The main parameters of known ULX pulsars: period $P$, period derivative $\dot {P}$, distance $d$, flux $F_{\rm x}$ (0.5-10 keV) and observation MJD. The $\checkmark$ marked calculation results are presented in Figure \ref{fig_spinup} - Figure \ref{fig_B_epm}}
	\label{tab:input}
	\begin{center}
		\begin{tabular}{l c c c c c l}
			\hline \hline
			source name & $P$ (s) & $\dot {P}$ (s/s)  & $d$ (Mpc) (G) & $F_{\rm x}$ (erg s$^{-1}$ cm$^{-2}$) & MJD &references \\
			\hline
			M82 X-2&	1.37&	-2.00E-10&	3.6&	4.18E-12&	56699&	\cite{Bachetti2014}\checkmark \\
			M82 X-2(spin down)&	1.38&	1.10E-10*&	3.6&	6.00E-13&	57641&	\cite{Bachetti2020} \\
			\hline
			ULX NGC 7793 P13&	0.4183891&	-4.03E-11*&	3.9&	2.70E-12&	57001&	\cite{Israel2017b} \\
			&	0.4169513&	-2E-12&	3.9	&5.19E-12&	57528&	\cite{Furst2016}\checkmark \\
			&	0.415669&	-6.00E-10&	3.9	&3.70E-12&	57924.11&	\cite{Furst2018} \\
			&	0.415214&	-1.00E-10&	3.9	&2.43E-12&	58083&	\cite{Furst2018} \\
			\hline
			ULX NGC 5907 ULX1&	1.427579&	-9.60E-09&	17.1&	3.14E-12&	52690.9&	\cite{Israel2017a} \\
			&	1.136042&	-4.70E-09&	17.1&	4.33E-12&	56851.5	&\cite{Israel2017a}\checkmark \\
			\hline
			NGC 300 ULX1&	126.28&	-1.148E-05&	1.88&	1.00E-12&	56978.6&	\cite{Vasilopoulos2018a} \\
			&	31.588&	-5.62E-07&	1.88&	4.60E-12&	57741&	\cite{Vasilopoulos2018a}\checkmark \\
			&	19.046&	-2.51E-07&	1.88&	3.16E-12&	58221&	\cite{Vasilopoulos2018a} \\
			\hline
			NGC 1313 X-2&	1.4576&	-1.38E-08&	4.2&	9.40E-12&	57998&	\cite{Sathyaprakash2019}\checkmark\\
			&	1.4612&	-3.20E-08&	4.2	&6.80E-12&	58096&	\cite{Sathyaprakash2019}\\
			\hline
			M51 ULX-7&	2.7977&	-2.4E-10&	8.58&	7.40E-13&	58283&	\cite{RodriguezCastillo2020}\checkmark\\
			\hline
			Swift J0243.6+6124&	9.854&	-6.80E-09&	0.007&	9.30E-09&	58031.7&	\cite{Doroshenko2018}\\
			&	9.846&	-1.75E-08&	0.007&	1.72E-07&	58057.3&	\cite{Doroshenko2018}\\
			&	9.827&	-2.22E-08&	0.007&	2.56E-07&	58067.1&	\cite{Doroshenko2018}\checkmark\\
			\hline
			SMC X-3&	7.8052&	-6.46E-09&	0.0621&	2.22E-09&	57624&	\cite{Weng2017}\checkmark\\
			&	7.7676&	-3.32E-10&	0.0621&	2.96E-11&	57763&	\cite{Weng2017}\\
			\hline \hline
		\end{tabular}
	\end{center}
\end{table*}

\begin{itemize}
	\item
	M82 X-2 is the first discovered ULX pulsar \cite{Bachetti2014}. Its period is around 1.37 s. This source is quite close to another ULX M82 X-1, which makes it difficult to accurately measure its flux. The only X-ray telescope that can be able to distinguish them is Chandra. The most interesting part of this source is that it has gone through the process of spin-up to spin-down in 2014, meaning that it is quite close to spin equilibrium \cite{Bachetti2020}.
	\item
	ULX NGC 7793 P13 was in its off-state during the first half year of 2020. However, unlike the last off-state which lasted for nearly two years during 2011-2012, at this time it only fainted for several months. In June 2020, it became active again \cite{Walton2020,Furst2020}. Nevertheless, the previously detected 0.417 s period \cite{Furst2016} was not found by XMM-Newton data \cite{Furst2020}.
	\item
	ULX NGC 5907 ULX1 is the brightest ULX pulsar discovered so far. Its luminosity can reach as high as $10^{41} \, {\rm erg/s}$. This source experienced two declines in X-ray luminosity in 2012-2014 and 2017-2018, respectively. The spin period is around 1.137 s \cite{Israel2017a}. The X-ray modulation was observed at around 78 days \cite{Walton2016}, which could be the orbital period.
	\item
	NGC 300 ULX1 is the only source with a very large spin period, from $\sim 126$ s at MJD 56978 \cite{Vasilopoulos2018a} to 31.6 s at MJD 57740 \cite{Carpano2018}, finally around 17 s \cite{Vasilopoulos2019}. The observed spin period showed the fast evolution with the period derivative of $10^{-7}-10^{-5}$ s s$^{-1}$. Another feature of this source is that its secular period evolution rate is quite stable, as well as its luminosity after MJD 57740 \cite{Carpano2018,Ray2019,Vasilopoulos2018a,Vasilopoulos2019}.
	\item
	NGC 1313 X-2 has very rare period data. Sathyaprakash et al. \cite{Sathyaprakash2019} found only two periods from six observations, and the measured period derivatives are negative as shown in Table~\ref{tab:input}. However, the calculated period derivative from the two observed periods is positive. They argued that this may be due to the affection of orbit period.
	\item
	The spin period of M51 ULX-7 is around 3.28 s at MJD 53552, and 2.798 s at MJD 58283 \cite{RodriguezCastillo2020}. The recent observed spin derivative is $\sim 1.5\times 10^{-10}$ s s$^{-1}$ \cite{RodriguezCastillo2020}.
	\item
	Swift J0243.6+6124 is the only detected ULX pulsar in the Milky Way. It underwent an outburst which started in October 2017, and reached its peak flux on November \cite{Wilson2018,Doroshenko2018,Wangp2020,Kong2020,Beri2021}. The spin period was around 9.84 s with a period derivative of $\sim 10^{-9}-10^{-8}$ s s$^{-1}$ \cite{Wilson2018}.
	\item
	SMC X-3 has not been widely studied as an ULX pulsar. There was a giant outburst happened on this source between July 2016 and March 2017. The peak Luminosity can reach $\sim 2.5\times10^{39} \,\rm erg/s$ \cite{Tsygankov2017}. The spin period is around 7.8 s and spin-up rate is correlated with its luminosity \cite{Weng2017}. However, Weng et al. \cite{Weng2017} also pointed out that the source may went though a spin-down to spin-up switch at the start of the outburst.
	
\end{itemize}

\subsection{Magnetic Fields Determined by Spin Period and Spin-Up Rate }

In this section, we will constrain the magnetic field and fastness parameter values using the six accretion torque models based on the observed parameters of these ULX pulsars. For a normal neutron star, we assume its mass, radius and moment of inertia are $M_{\rm *}=1.4 M_{\rm \odot}, R_{\rm *}=10 \, \rm km$, $I = 10^{45} \,\rm g \, cm^2$, respectively. These values are set only for calculation convenience. The uncertainties in these values (e.g., $M_{\rm *}=0.9 M_{\rm \odot}$ or $2 M_{\rm \odot}$ as taken in \cite{Erkut2020}) may change the field results by around a factor of ten, depending on different models. $\delta$ has the range of 0.01 to 0.3 \cite{Erkut2017}, and $b$ ranges in 0.01-1 \cite{Erkut2020,Tong2019}. When considering the condition that $R_{\rm in}>R_{\rm *}$, according to the definition of $\omega_{\rm *}$ \cite{Erkut2004}, a minimum limit is set as $(R_{\rm *}/R_{\rm co})^{3/2}$. We should note that, normally, we adopt the assumption that $\omega_{\rm *} \leq1$.

In LRB model, the mid-plane temperature of the accretion disk is assumed to be $10^5 \, \rm K$, thus according to the relation $c_{\rm s}\equiv (k_{\rm BT}/m_{\rm p})^{1/2}$, Newton sound speed can be derived. The thermal pressure parameter $p_{\rm 0}$ in WangC model is set as $10^6 \, \rm dyn$.

\begin{figure*}
	\centering
	\includegraphics[width=0.95\textwidth]{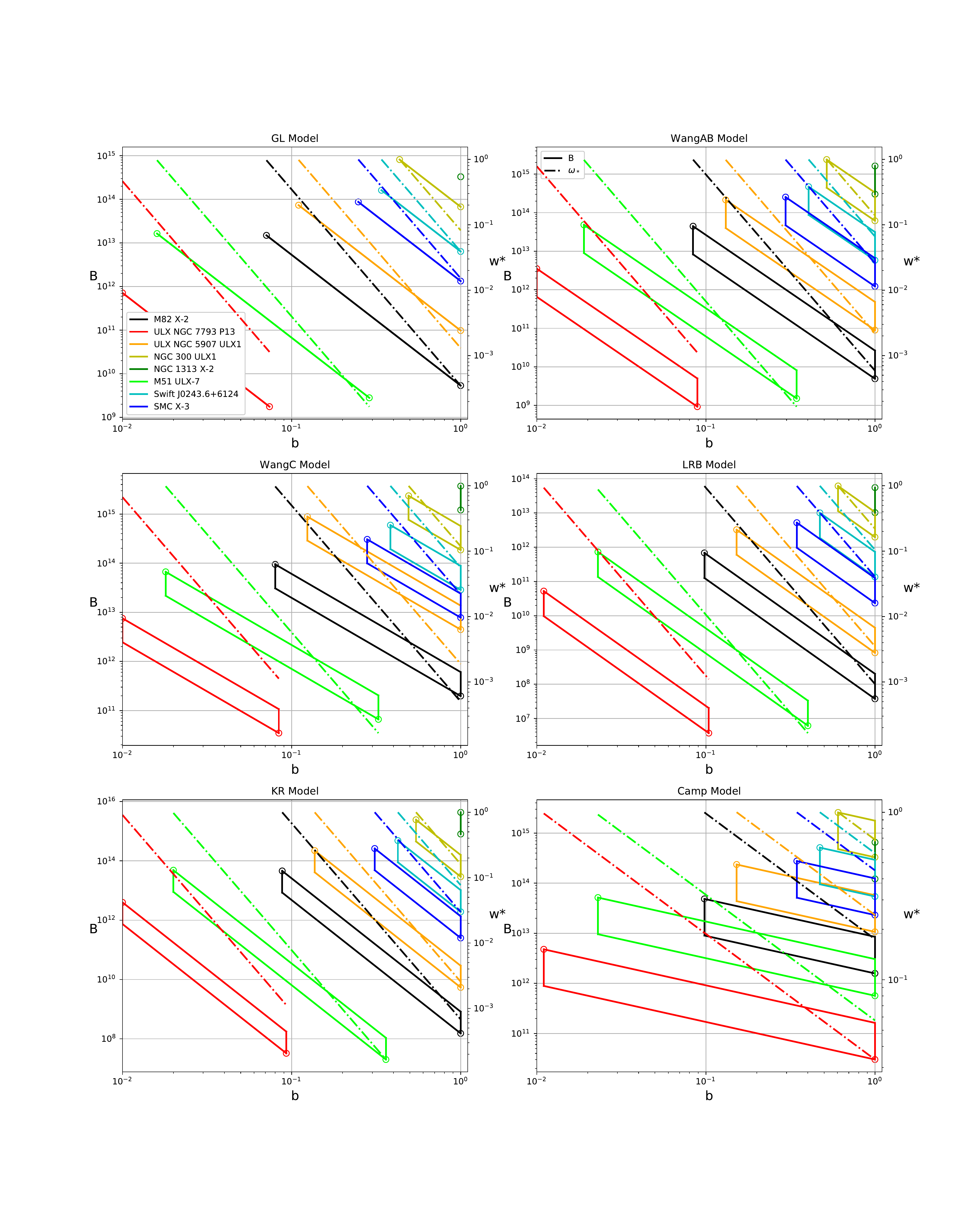}
	\caption{Fastness parameter and dipole magnetic field versus beaming factor $b$ in different torque models. Assuming the sources are away from spin equilibrium. As shown in the legend, each color represents a different source. The dash-dotted lines and the solid lines are the ranged for fastness parameters and magnetic fields, respectively. The maximum and minimum values of each source are marked as open circles. }
	\label{fig_spinup}
\end{figure*}

\begin{figure*}
	\centering
	\begin{minipage}{0.45\textwidth}
		\includegraphics[width=0.95\textwidth]{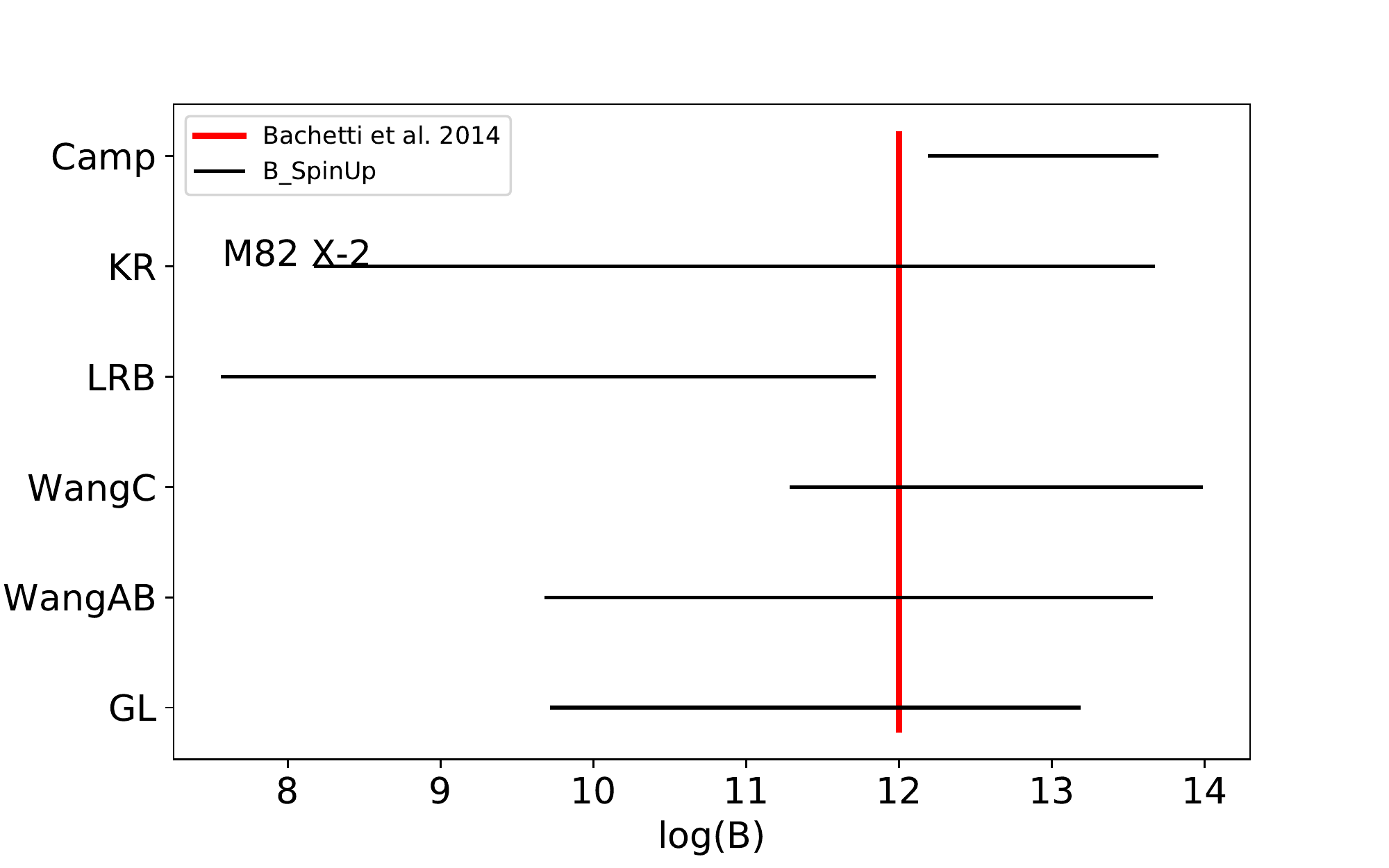}
	\end{minipage}
	\begin{minipage}{0.45\textwidth}
		\includegraphics[width=0.95\textwidth]{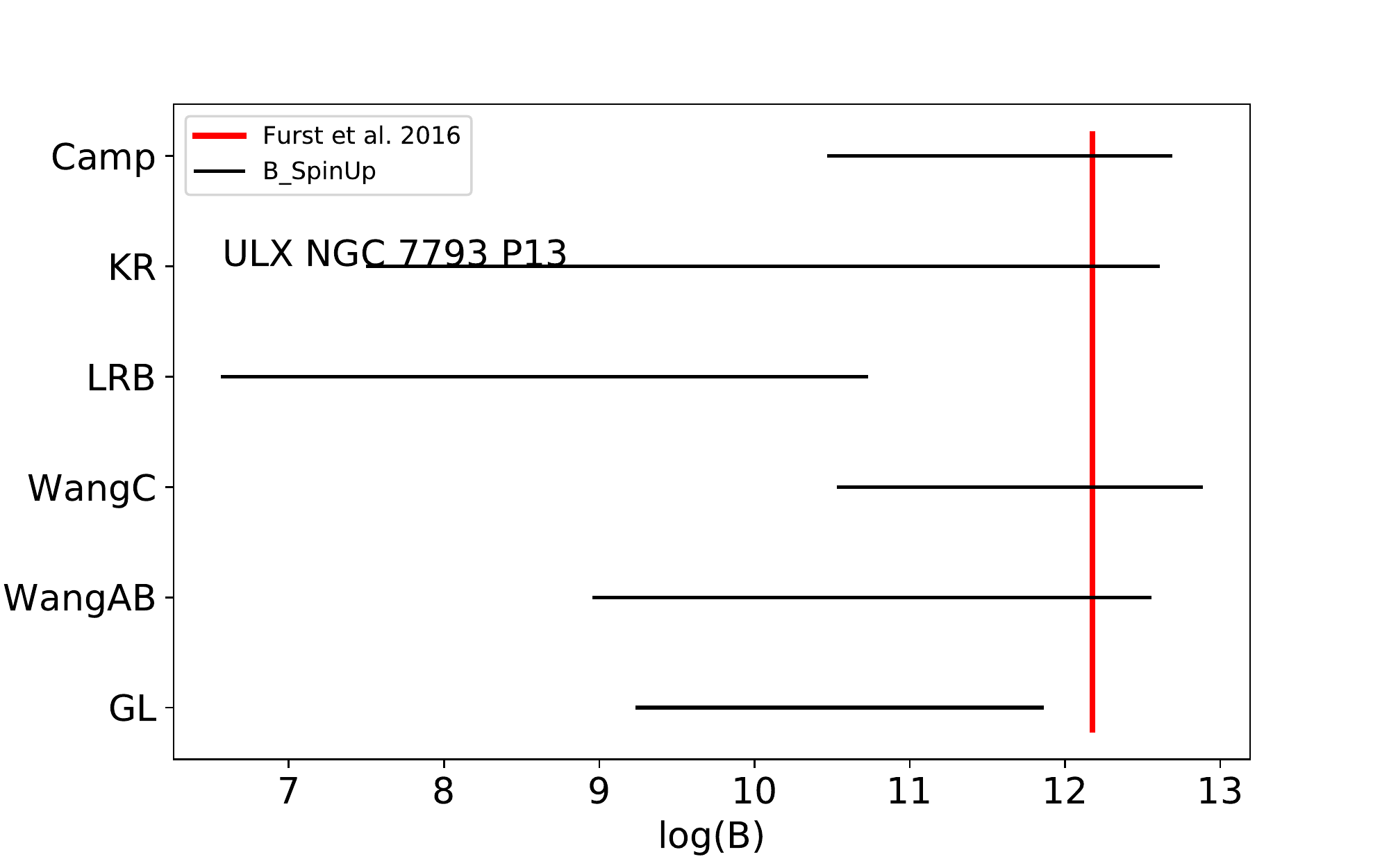}
	\end{minipage}
	\begin{minipage}{0.45\textwidth}
		\includegraphics[width=0.95\textwidth]{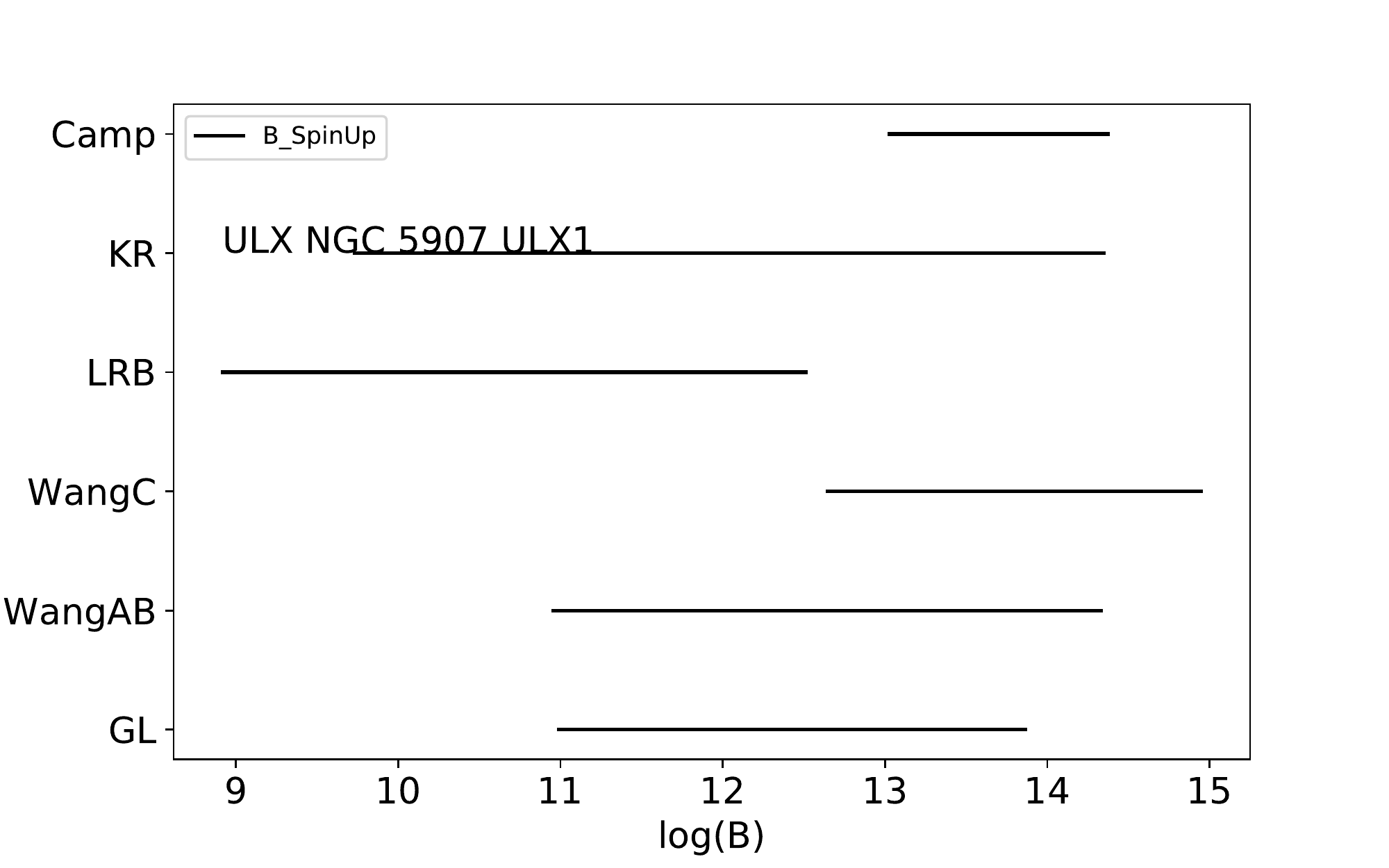}
	\end{minipage}
	\begin{minipage}{0.45\textwidth}
		\includegraphics[width=0.95\textwidth]{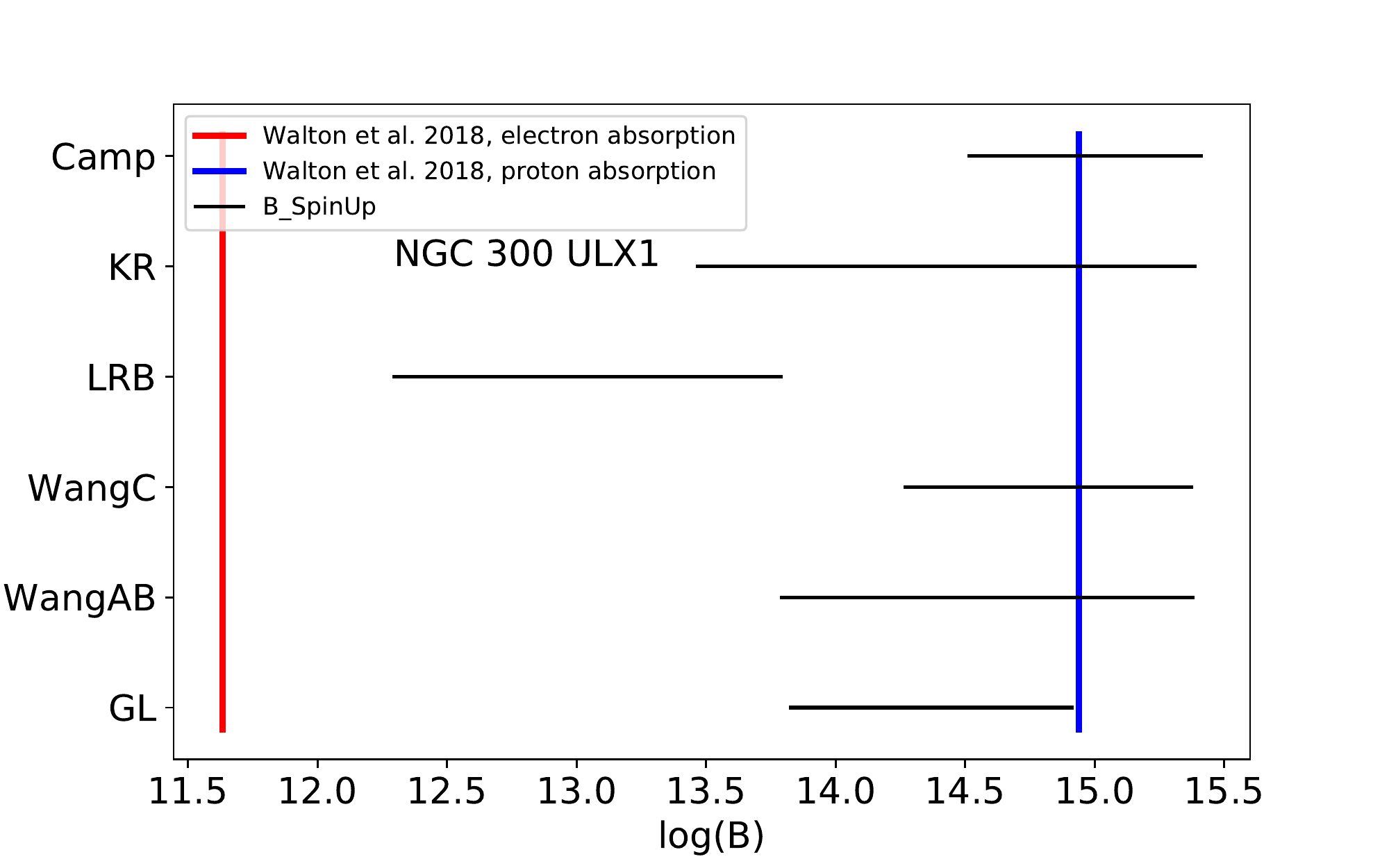}
	\end{minipage}
	\begin{minipage}{0.45\textwidth}
		\includegraphics[width=0.95\textwidth]{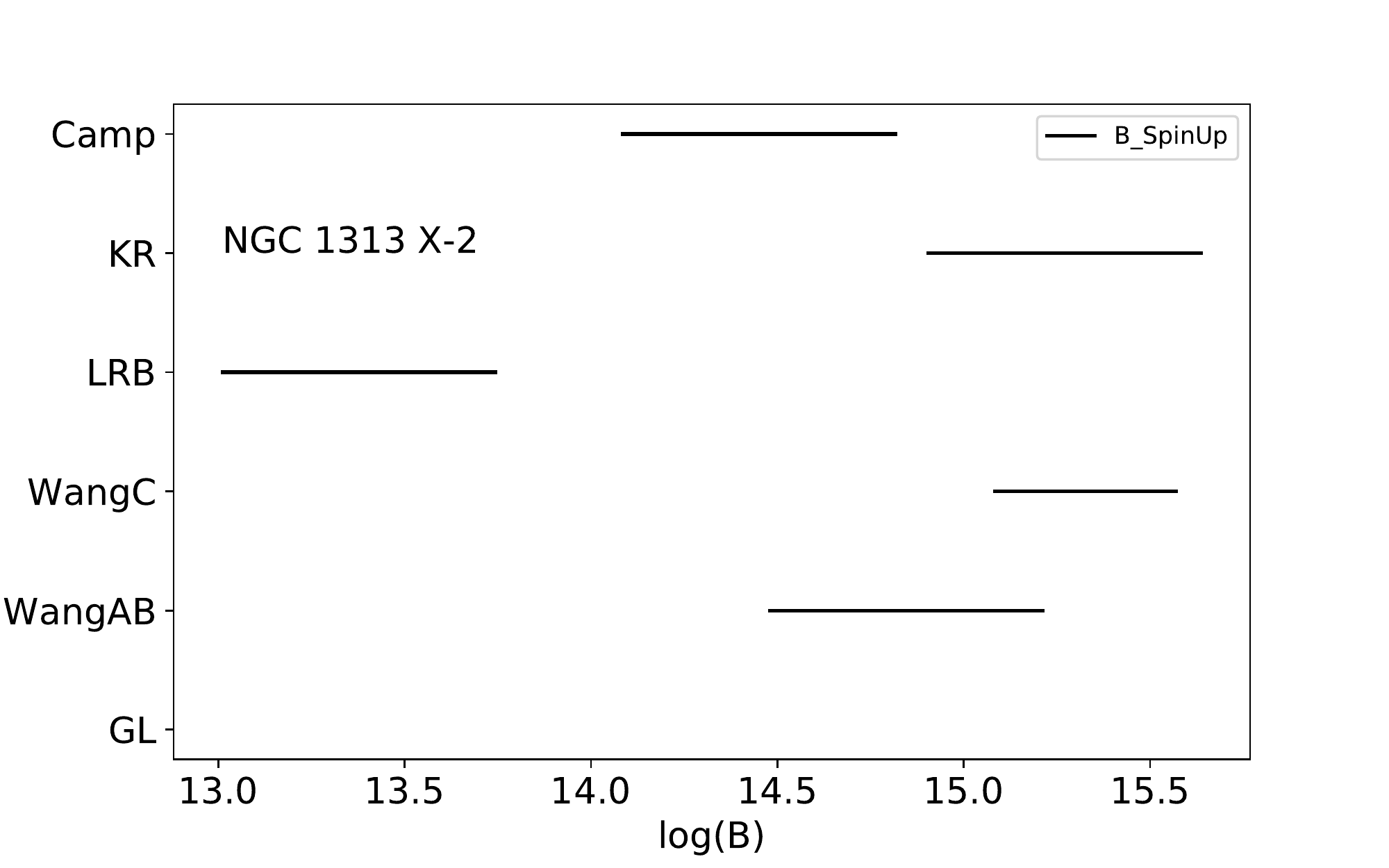}
	\end{minipage}
	\begin{minipage}{0.45\textwidth}
		\includegraphics[width=0.95\textwidth]{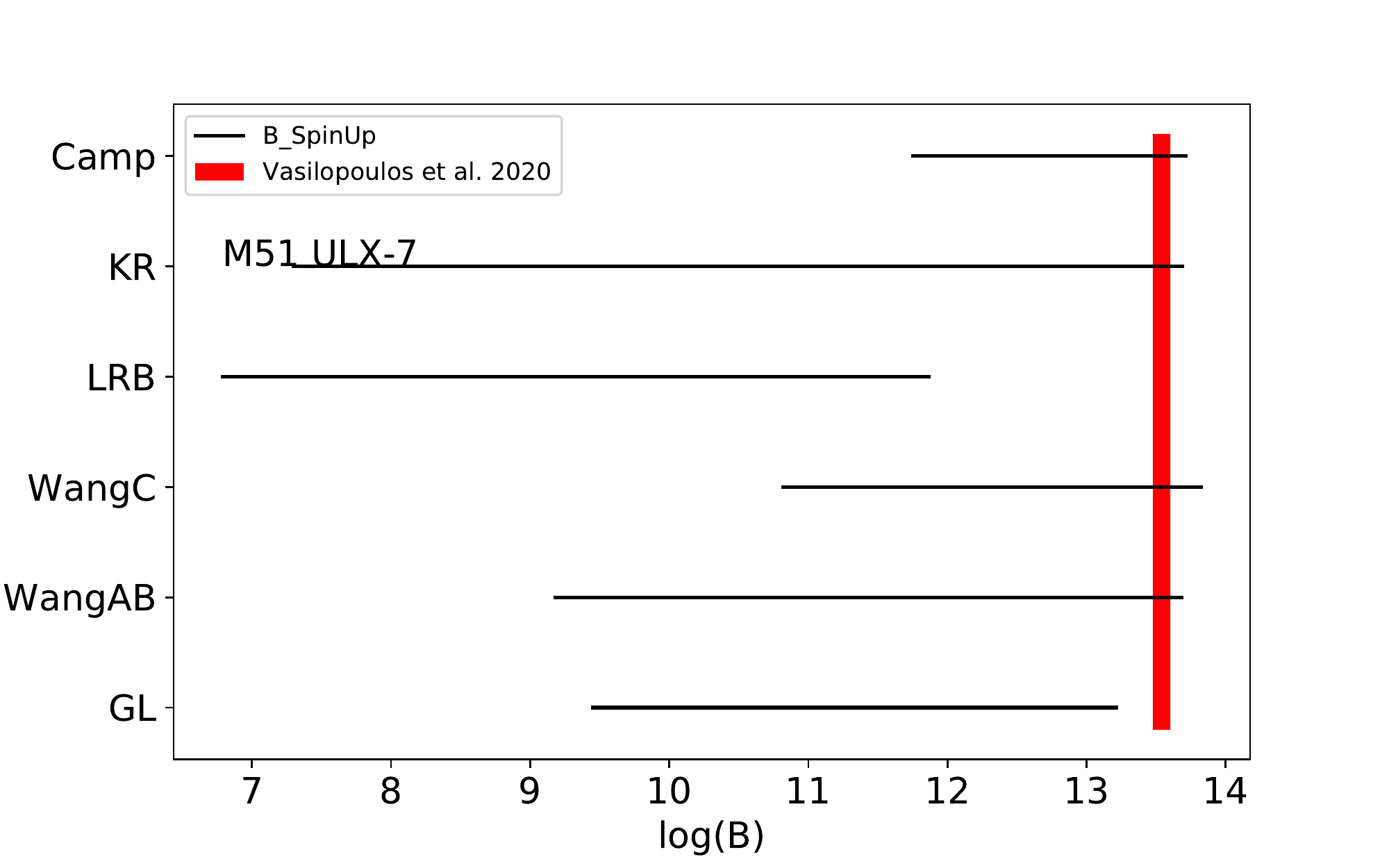}
	\end{minipage}
	\begin{minipage}{0.45\textwidth}
		\includegraphics[width=0.95\textwidth]{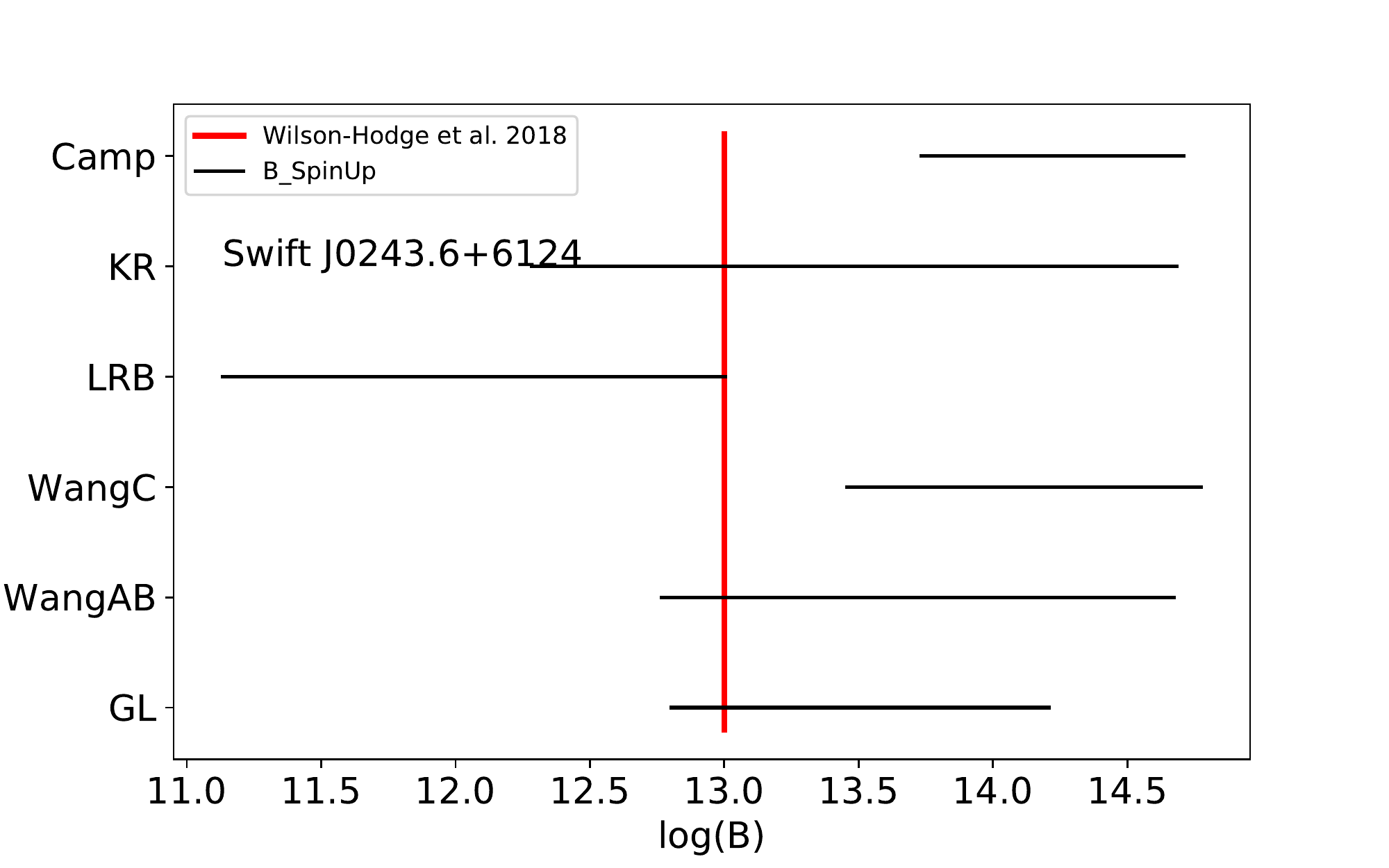}
	\end{minipage}
	\begin{minipage}{0.45\textwidth}
		\includegraphics[width=0.95\textwidth]{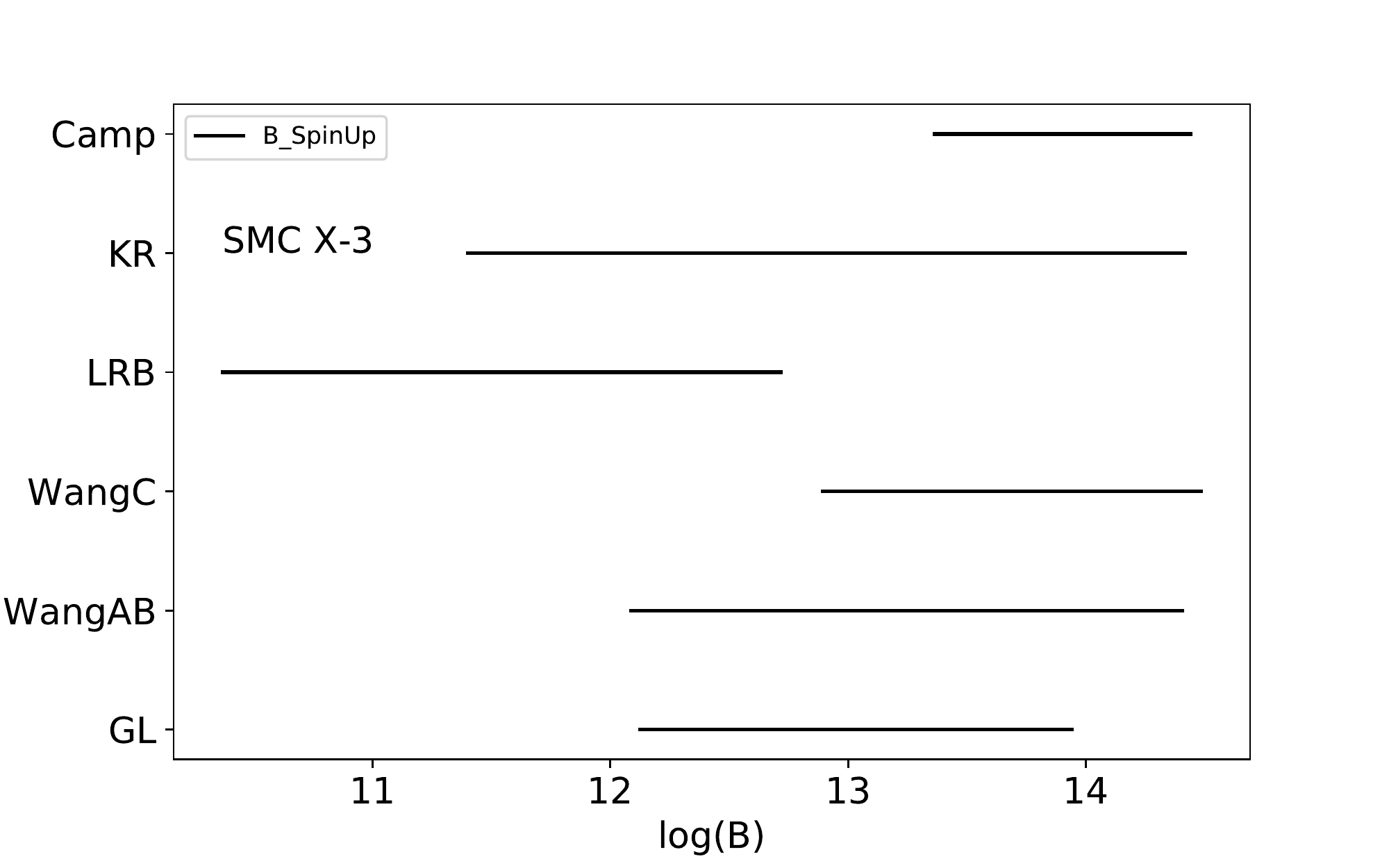}
	\end{minipage}
	\caption{The magnetic field values constrained by different torque models plotted in logarithms assuming the spin-up regime for each source. The literature fields are also presented in red and blue lines/ranges, including the possible cyclotron absorption line determined fields in NGC 300 ULX1, the modulation to the precession determined fields in M51 ULX-7, plus the spin-up determined values in M82 X-2, ULC NGC 7793 P13 and Swift J0243.6+6124. }
	\label{fig_B}
\end{figure*}

As noted above, the $*$ marked data sets in Table~\ref{tab:input} are not used in $B$ calculation. Moreover, we try to avoid showing the results where all the $\omega_{\rm *}$ are larger than 1, because this situation will cause some errors. Thus, the data sets that meets the requirements for each source are selected. For M82 X-2, ULX NGC 7793 P13, NGC 300 ULX1, M51 ULX-7 and SMC X-3, only one set of data satisfies the conditions. Two data sets of Swift J0243.6+6124 can be chosen, and the differences in magnetic fields are quite small, around 0.1 in logarithm. The results of ULX NGC 5907 ULX1 are both well: the minimum magnetic field of the first data set is around one order of magnitude higher than the second one for each model, while the maximum values are consistent with each other. NGC 1313 X-2, however, has no solution with $\omega_{\rm *}\leq1$, even though we take its neutron star mass as low as $1.2 M_\odot$ \cite{Stair2004}. To keep the results self-consistent, we still adopt $1.4 M_\odot$ in the calculation, and use the magnetic field results of $b=1$, which will be also discussed in the next section. We have selected derived plots for each source of typical parameters for the presentation and estimating magnetic field ranges, and the chosen data set for each source is $\checkmark$ marked in Table~\ref{tab:input}. In Figure \ref{fig_spinup}, we determined fastness parameters and dipole magnetic fields of eight sources in various accretion torque models considering the spin-up rate. To compare the magnetic field values derived by different models, we showed the magnetic field ranges for each source in each panel of Figure \ref{fig_B}.

For the spin up regimes, the derived magnetic field values are located in the range of $10^9$ to $10^{15} \, \rm G$. The GL, WangAB, WangC and Camp models predict the similar magnetic field value ranges for each source, e.g., for M82 X-2 , ULX NGC 7793 P13 and M51 ULX-7, $B\sim 10^{11-13}$ G, for NGC 300 ULX1, NGC 5907 ULX1, NGC 1313 X-2, Swift J0243.6+6124 and SMC X-3, $B\sim 10^{13-15}$ G.  Generally, WangC model gives the largest maximum magnetic fields, but are very close to Camp, KR and WangAB maximum values. The maximum and minimum values of the LRB model are the smallest, and can be as low as $ \sim 10^7 \, \rm G$. In the meanwhile, its maximum and minimum interval span is quite large, as well as KR model. For the case of the spin-up regimes, the calculated results of magnetic field value ranges are sensitive to $\dot P$. Since spin period measurements may be affected by orbital period of the binary system, and the very rare data at present and really small values of $\dot{P}$ of some ULX pulsars bring uncertainties on the calculations and constraints on the magnetic fields.

\subsection{Magnetic Fields Determined by Spin Equilibrium}
\label{section_spineqm}
\begin{figure*}
	\centering
	\includegraphics[width=0.95\textwidth]{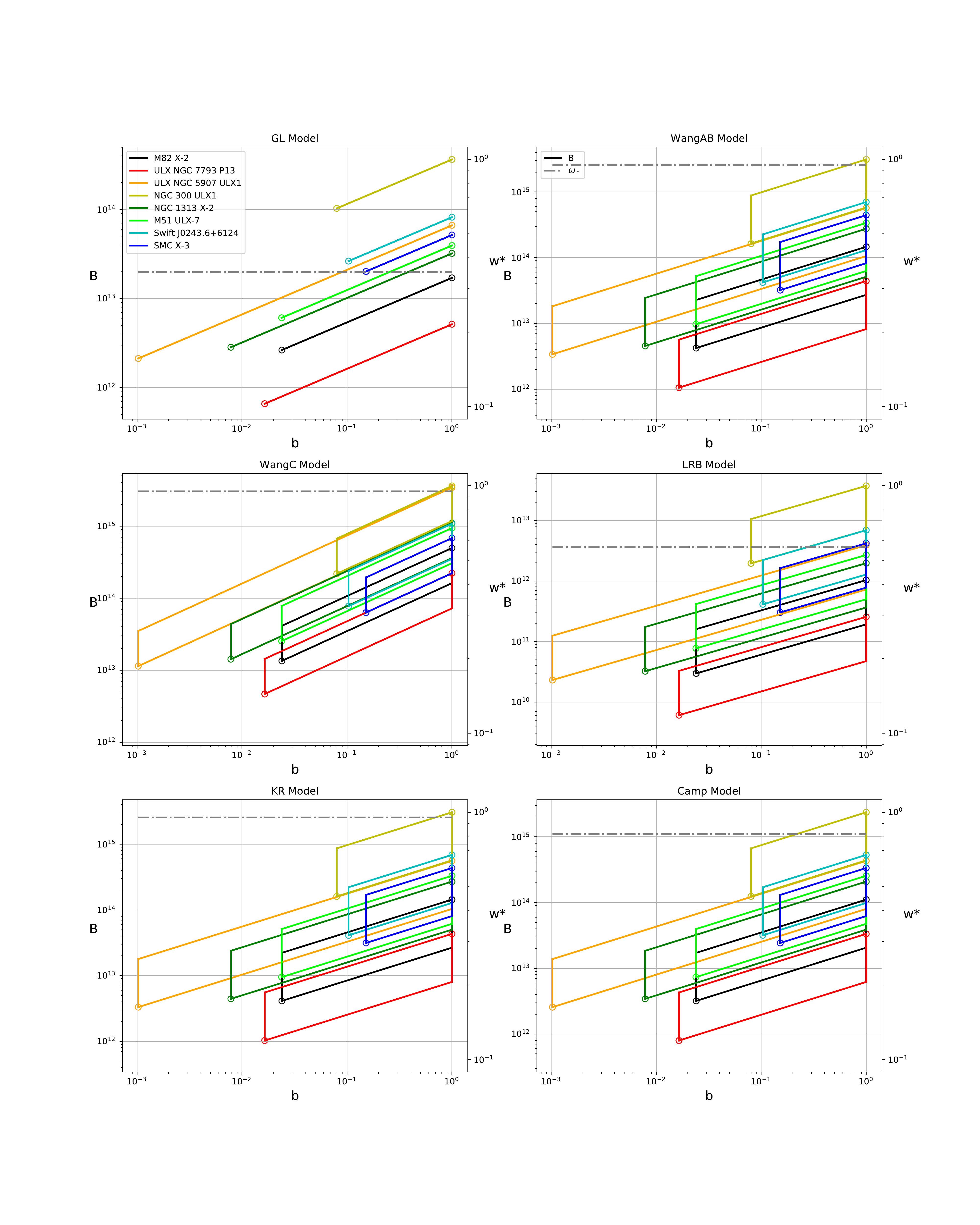}
	\caption{Fastness parameter and dipole magnetic field versus $b$ assuming the sources are near equilibrium. As shown in the legend, each color represents a different source. The gray dash-dotted lines and the solid lines are critical fastness parameters and magnetic fields, respectively. The maximum and minimum of each source are marked as open circles.}
	\label{fig_eqm}
\end{figure*}

\begin{figure*}
	\centering
	\begin{minipage}{0.45\textwidth}
		\includegraphics[width=0.95\textwidth]{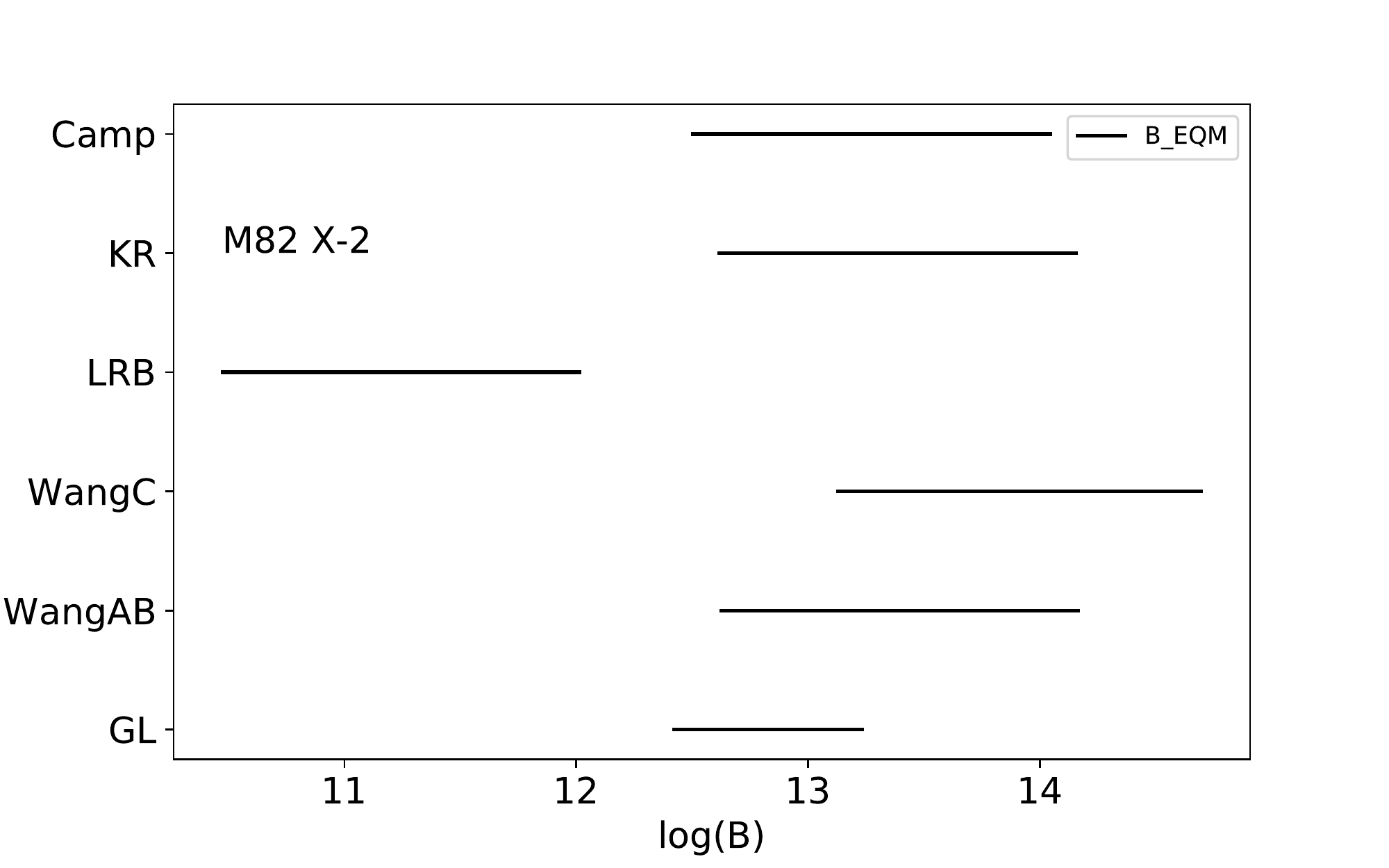}
	\end{minipage}
	\begin{minipage}{0.45\textwidth}
		\includegraphics[width=0.95\textwidth]{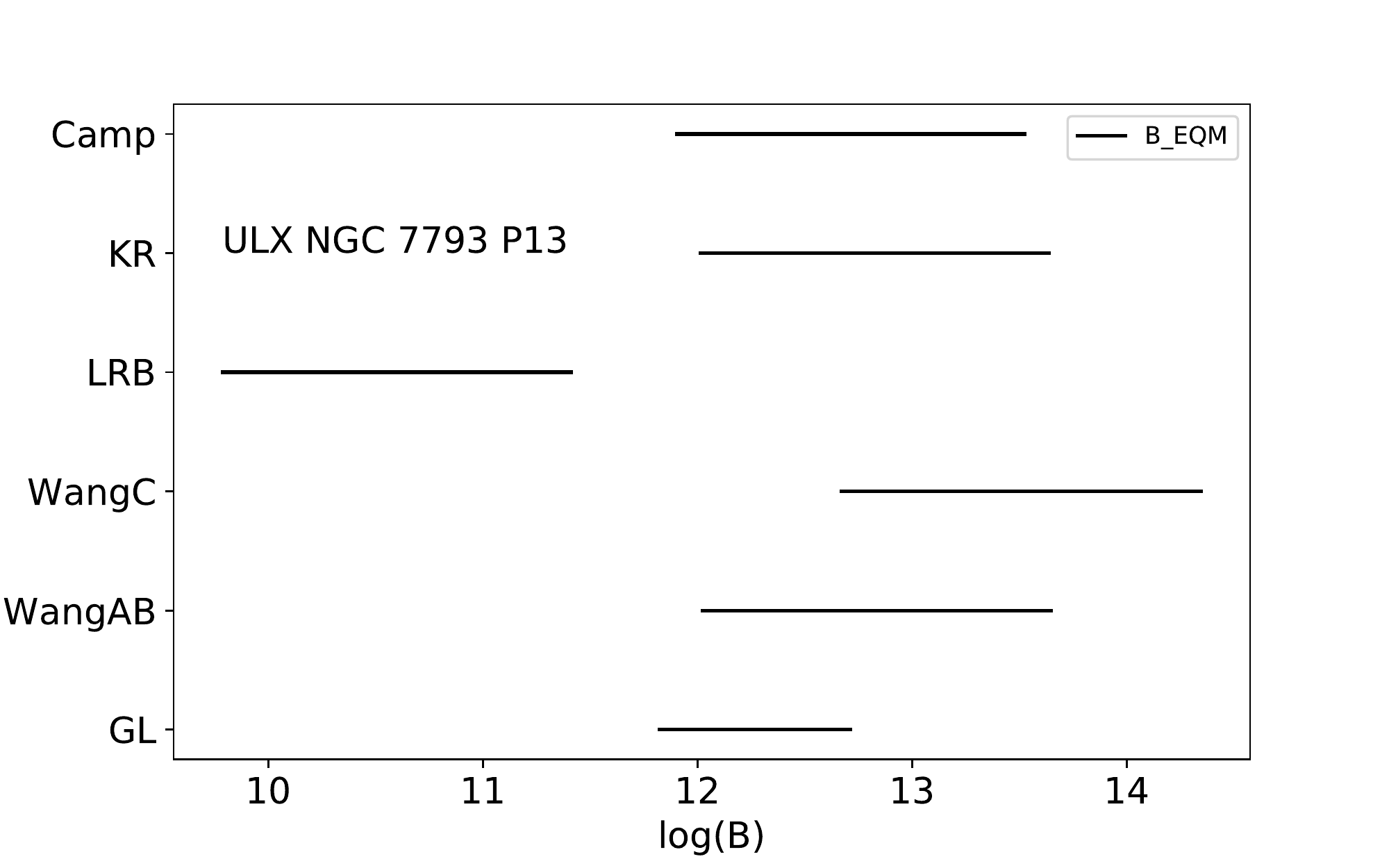}
	\end{minipage}
	\begin{minipage}{0.45\textwidth}
		\includegraphics[width=0.95\textwidth]{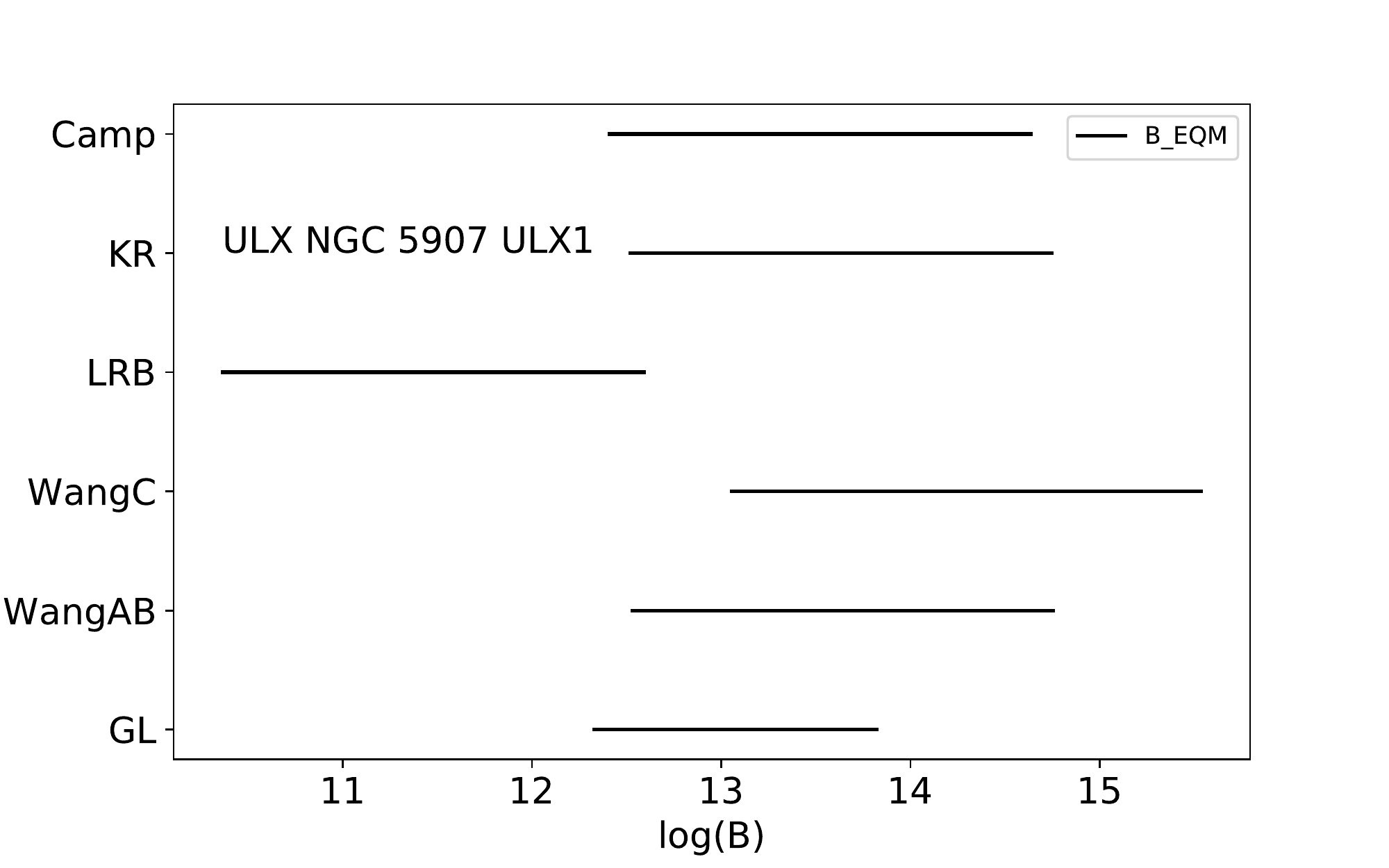}
	\end{minipage}
	\begin{minipage}{0.45\textwidth}
		\includegraphics[width=0.95\textwidth]{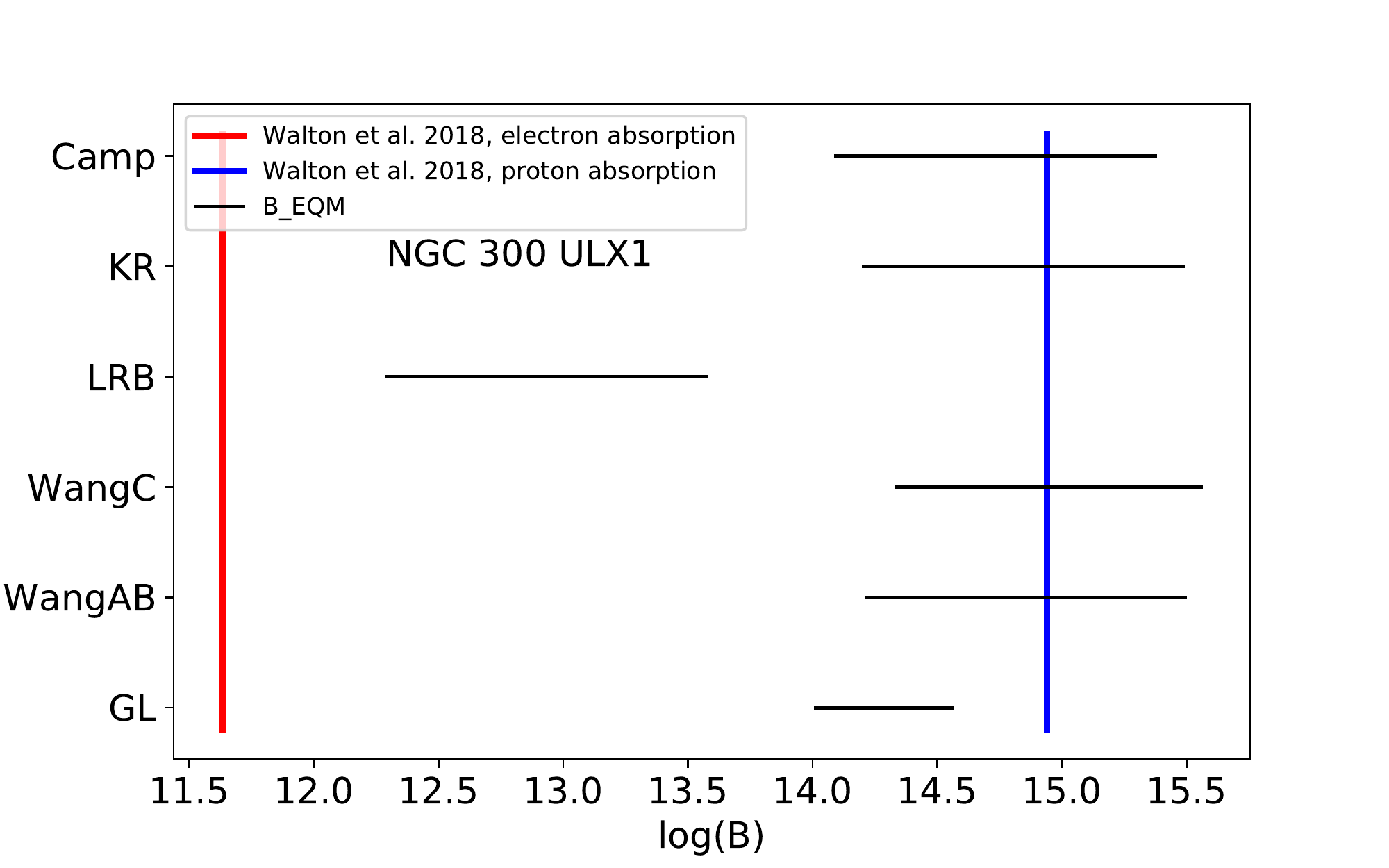}
	\end{minipage}
	\begin{minipage}{0.45\textwidth}
		\includegraphics[width=0.95\textwidth]{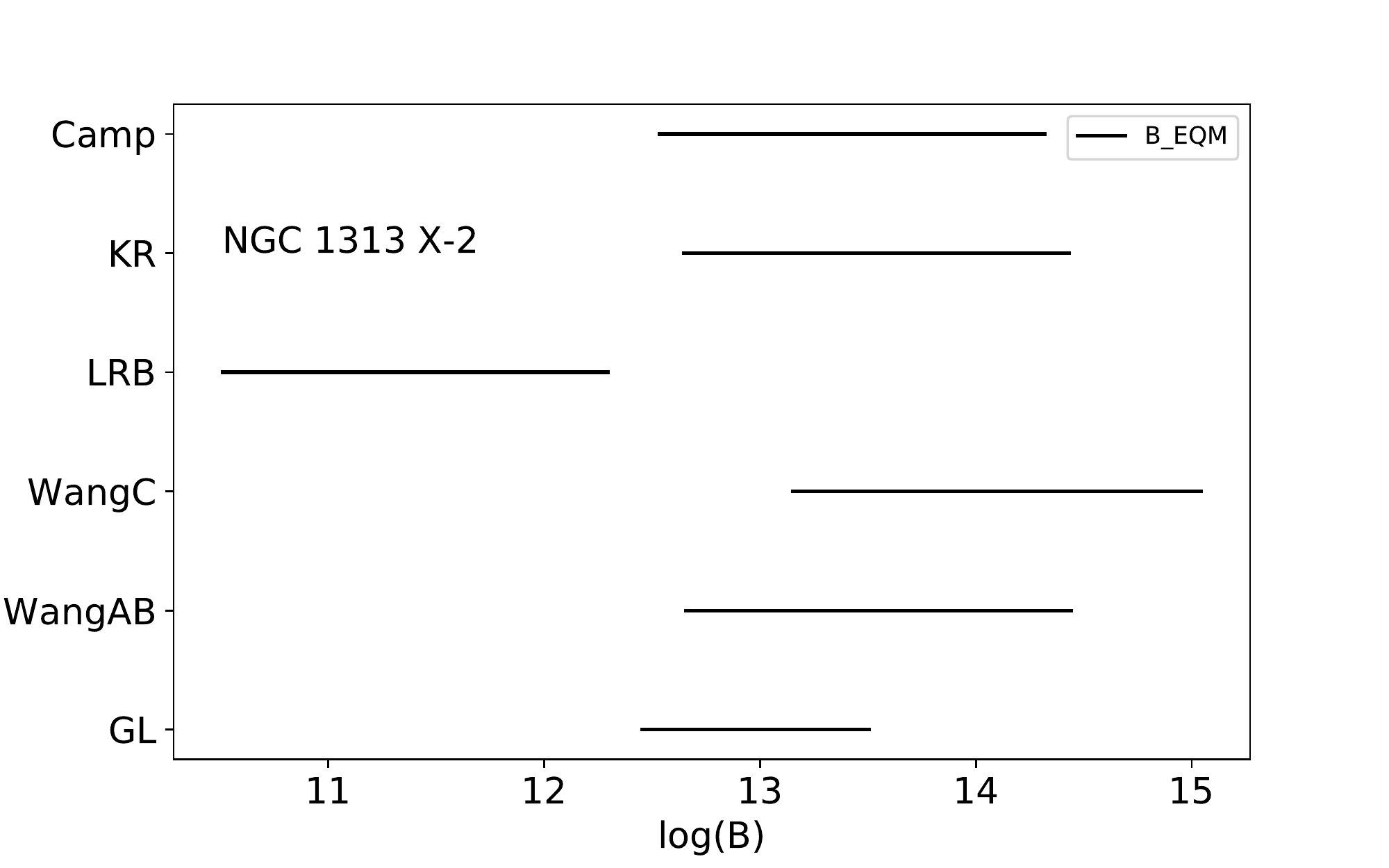}
	\end{minipage}
	\begin{minipage}{0.45\textwidth}
		\includegraphics[width=0.95\textwidth]{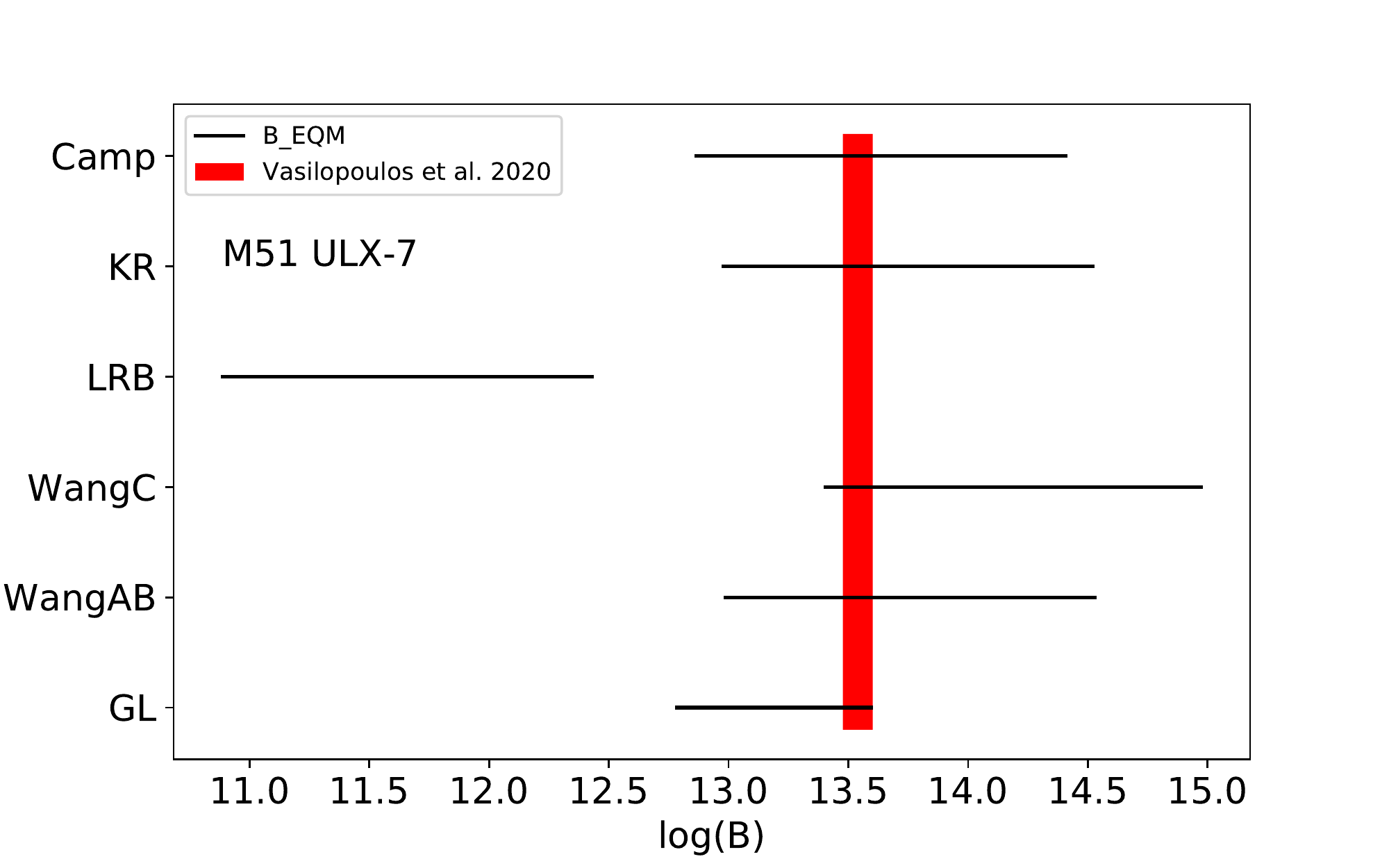}
	\end{minipage}
	\begin{minipage}{0.45\textwidth}
		\includegraphics[width=0.95\textwidth]{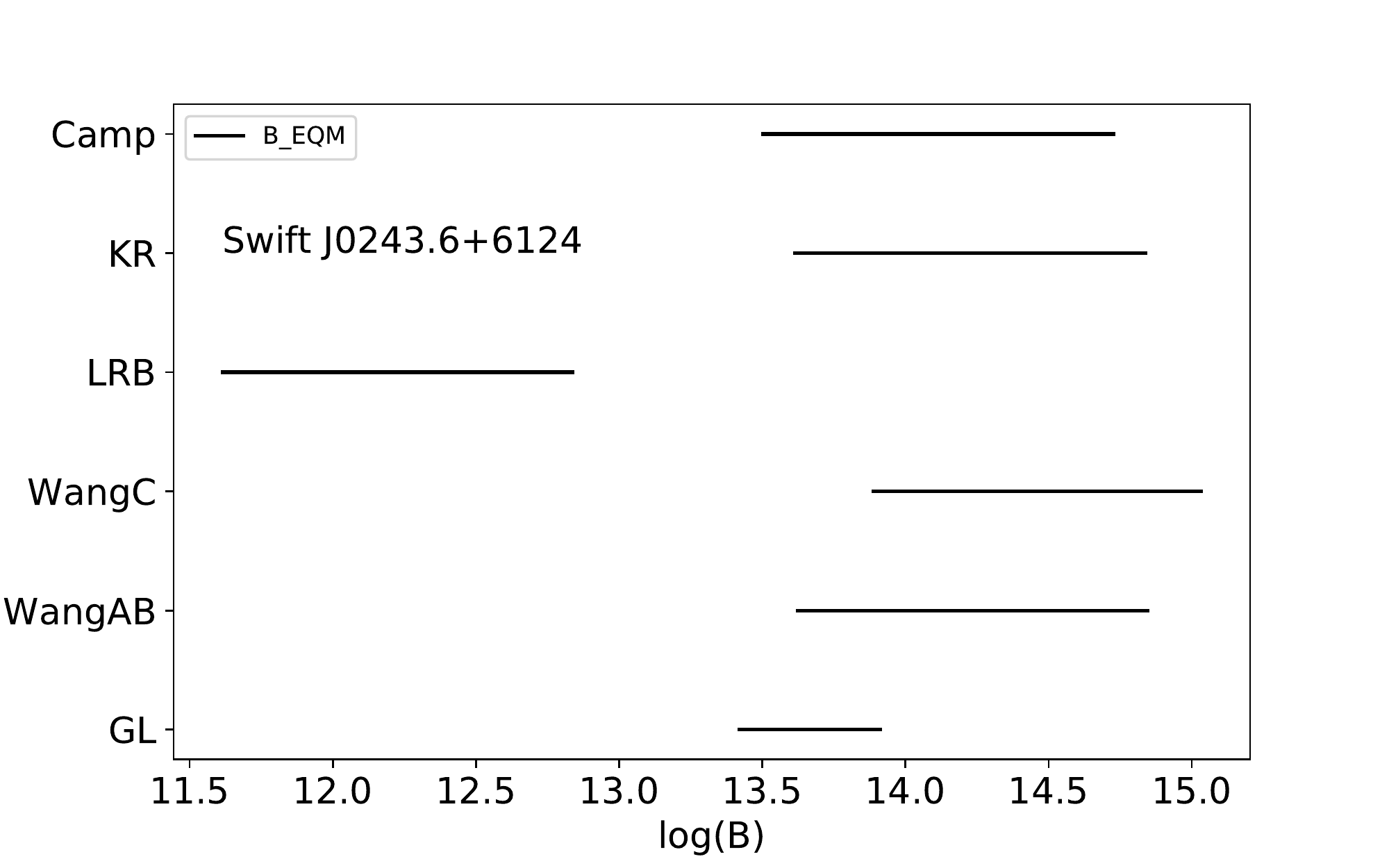}
	\end{minipage}
	\begin{minipage}{0.45\textwidth}
		\includegraphics[width=0.95\textwidth]{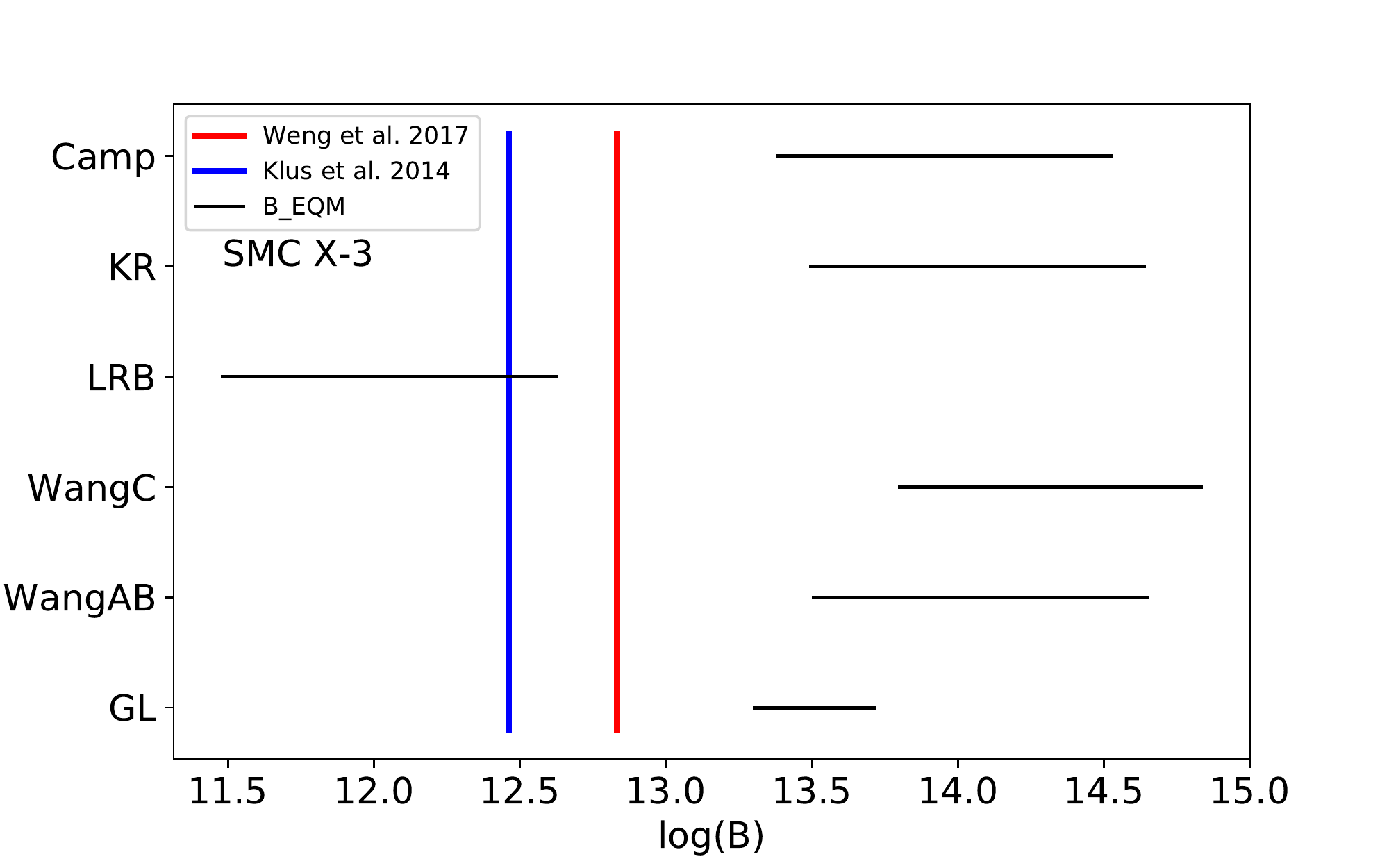}
	\end{minipage}
	\caption{The magnetic fields constrained by different torque models plotted in logarithms assuming the spin equilibrium. The literature fields are also presented in red and blue lines/ranges, including the possible cyclotron absorption line determined fields in NGC 300 ULX1, the modulation to the precession determined fields in M51 ULX-7, and the spin equilibrium determined values in SMC X-3.}
	\label{fig_B_epm}
\end{figure*}

There are evidences showing that ULX pulsars may be very close to spin equilibrium state, such as the sudden convert of spin derivative sign of M82 X-2 and the short spin change time scales \cite{Bachetti2014,Israel2017a,Israel2017b,Furst2016,Bachetti2020}. We again adopt the range of $\delta$ from 0.01 to 0.3, and $b$ in the range of $0.01-1$ \cite{Erkut2020,Tong2019}. We substitute these values into the derived equilibrium magnetic field equations to constrain the magnetic field values for each ULX pulsar in the equilibrium state. Figure \ref{fig_eqm} shows the calculated results for all sources. Each panel of the Figure \ref{fig_eqm} represents one accretion torque model. We collect the magnetic field values for each source constrained by six models for comparison, which are presented in Figure \ref{fig_B_epm}.

As for the spin equilibrium regime, the LRB model always predicts the lower magnetic field ranges of $B\sim 10^{10-12}$ G for all sources. The other four torque models predict the magnetic field values of $10^{12-13}$ G for ULX NGC 7793 P13, while WangC model is slightly higher around $10^{13-14}$ G. For other seven ULX pulsars, the predicted magnetic field values from five models distributed analogously in the range of $10^{13-15} \,\rm G$ which are the typical values for magnetars.

\section{Discussions \& Conclusions}\label{sec:4}
Generally, we assumed that ULX pulsars should be the accreting magnetars, producing the super-Eddington pulsed luminosities. However, up to now, the real magnetic field values in these systems are still in dispute. In this work, we presented the magnetic field values of eight ULX pulsars using six different accretion torque models. Science aims here are to check the effects of magnetic field constraints on ULX pulsars by different accretion torque models. Though the limited number of ULX pulsars greatly restricts our understanding of their physical process at present, we use the currently available flux and period data, combined with different accretion torque models, to analyze the magnetic field, accretion process and rotational behavior of the neutron stars. Both away-from-equilibrium (spin-up process) and near-equilibrium regimes are considered and discussed.

Assuming the spin-up regimes for all ULX pulsars, the torque models of WangAB/C, Camp, GL generally predict a stronger magnetic field, approaching the ranges of magnetars. The KR model gives a wide range of the magnetic field values, still consistent with the predictions by the Wang, Camp, GL models. However, the LRB model generally predict a lower value range for the ULX pulsar magnetic field (as low as $\sim 10^{8-12}$ G). For NGC 1313 X-2, all accretion torque models suggest a strongly magnetized neutron star with $B> 10^{13}$ G.

For the case of the spin equilibrium, the WangAB/C, GL, Camp and KR models have the similar prediction on the magnetic fields for all ULX pulsars, $B\sim 10^{13-14}$ G. The LRB model still gives a lower magnetic field values for all sources, $B\sim 10^{12}$ G. Thus the LRB model may be more suitable for the normal accretion neutron star systems, and other torque models would prefer to an accretion magnetar scenario.

As mentioned above, the LRB model always predicts a smaller value of magnetic field, which may have two possibilities. Firstly, as argued by Lovelace et al. \cite{Lovelace1995}, the magnetic form of the outer part of the disc in their model is not correct, and thus suggest the generation of open lines and outflows. Then they got the turnover radius, $R_{\rm to}$, by least-squares fitting, and derived the magnetic field using $R_{\rm to}$ in the spin-up regime. However, we substituted the observation data into their equation and got an unlikely small field value. Thus the open lines and outflows hypothesis may not be suitable for the magnetic field calculation in ULX pulsars. The other probability is the Newton sound speed we utilized is much lower than it should be, which means a higher mid-plane temperature. The accretion disks of ULX pulsars may be much hotter than we assume. If the LRB model need to give a similar prediction to other torque models, the disc temperature should be $\sim10^{12} \,\rm K$ which leads to $c_{\rm s} \sim 9\times10^9 \,\rm cm/s$, which is unphysical in ULX pulsars.

Previous literatures also estimated the magnetic field values for these ULX pulsars. The calculations were based on the torque models in the spin-up regimes \cite{Furst2016,Ray2019,Vasilopoulos2020,RodriguezCastillo2020,Wilson2018,Zhang2019,Kong2020}, or the spin equilibrium assumption (for SMC X-3 \cite{Klus2014,Weng2017,Tsygankov2017}), or the critical luminosity \cite{Erkut2020}. These theoretical results have also predicted the wide value ranges of the magnetic field for ULX pulsars. Our calculations are still comparable to previous work, and we considered its possible dependence on the different accretion torque models. The better way to constrain the magnetic field of ULX pulsars may require more independent observations.

The cyclotron resonant scattering features are the unique way to directly measure the magnetic field in accreting X-ray pulsars. It is interesting that a possible cyclotron absorption line at $E\sim 13$ keV was reported in NGC 300 ULX1 \cite{Walton2018}. If this feature was real, we could derive the surface magnetic field of the neutron star in NGC 300 ULX1 around $1\times10^{12}/(1+z_g)$ for the electron absorption, and $2\times10^{15}/(1+z_g)$ if CRSF is caused by proton absorption, where $z_g\sim 1.3$ is the gravitational redshift near the surface of neutron stars. In both the spin-up regime and near-equilibrium case, the GL, Wang, KL and Camp models all predict $B\sim 10^{14-15} \,\rm G$ for NGC 300 ULX1, which may be consistent with the proton absorption line case. However, the broadness of the line in NGC 300 ULX-1 may not support this result, since proton lines are often narrow \cite{Brightman2018}. A possible explanation is that the assumed mass and radius of the neutron star utilized for this source is not appropriate, and the variances in the observed parameters in Table~\ref{tab:input} may also cause some errors. In the meanwhile, the LRB model gives lower magnetic field values for NGC 300 ULX1: $B\sim (2-62)\times 10^{12}$ G for the spin-up regime; $B\sim (2-37)\times 10^{12}$ G for the near-equilibrium case. If the observed CRSF is due to the electron absorption, the LRB model may explain the observed parameters of NGC 300 ULX1.

Vasilopoulos et al. \cite{Vasilopoulos2020} reported a super-orbital modulation period at $\sim 39$ days for M51 ULX-7 using the Swift/XRT lightcurves. They have argued that the super-orbital periodicity is due to disc precession, which is attributed to the free precession of the neutron star. With simple assumptions, they estimated a magnetic field of $B\sim (3-4)\times 10^{13}$ G. Though this independent method has uncertainties, we can still compare our estimations with it. As shown in Figure~\ref{fig_B}, in the spin-up regimes, all accretion torque models predict magnetic fields lower than this range. In the near equilibrium situation in Figure~\ref{fig_B_epm}, the WangAB/C, GL, KR and Camp models predicted a magnetic field range of $B\sim 10^{13}-10^{14}$ G which is consistent with the value by the modulation period \cite{Vasilopoulos2020}. From these independent constraints on magnetic field by different methods, M51 ULX-7 would be near the spin equilibrium.

The special case M82 X-2 underwent the spin-up to spin-down convert in 2014 \cite{Bachetti2014,Bachetti2020}, suggesting that this source is located near the spin equilibrium. In our calculations, the WangAB/C, GL, Camp and KR models implied the magnetic field range of $4\times 10^{12}-5\times10^{14}$ G, and the LRB model gave a range of $10^{11-12}$ G. This source is the only one which showed the spin-down process. So during the spin-down process, we suggest that the spin-up torque from the accretion flows is smaller than the spin-down torque by the magnetic torque. Then we can predict a lower limit of the magnetic field of the neutron star in M82 X-2. The magnetic torque should be larger than the observed spin-down torque of the system, i.e., $ \mu^2/R_{\rm co}^3 \geq I|\dot \omega|$, where $\dot \omega$ is the spin-down rate. The average spin-down rate of M82 X-2 between 2014 and 2016 was $\dot P \sim1.1\times 10^{-10}$ s s$^{-1}$, we can derive a magnetic field limit for M82 X-2 that $B>3\times 10^{12}$ G. From Figures~\ref{fig_B} \& \ref{fig_B_epm}, the magnetic field ranges by the LRB model are lower than $10^{12}$ G for both the spin-up regime and equilibrium case. The other torque models predict the consistent magnetic field ranges, then can be applicable to the source M82 X-2.

NGC 1313 X-2 have the high magnetic field values of $B> 10^{13}$ G based on the different torque models. Thus, it should be a good candidate for accreting magnetars \cite{Wang2013,Tong2019}. In the calculations, we have taken the neutron star mass as 1.4 $M_\odot$. The lower neutron star masses would lead to lower magnetic field values (e.g., Erkut et al. \cite{Erkut2020} assumed the neutron star mass to be 0.9 $M_{\rm \odot}$ to reduce the magnetic field for ULX pulsars). However, the measured mass range of the neutron star in X-ray binaries in our Galaxy is $\sim 1.2-2 M_\odot$ \cite{Stair2004}. Thus the low mass of neutron star at $\sim 0.9 M_\odot$ may be not reasonable from the present observations. As mentioned in Section 3.2, using 1.2 $M_\odot$ in calculations will also give no solution with $\omega_{\rm *}\leq1$. By adopting field values for $b=1$, we still have $B > 10^{14}$ except the LRB model which predicts $4\times10^{12} < B < 2\times10^{13}$. For comparison, if taking 0.9 $M_\odot$ as the stellar mass, we obtain $7\times10^{11} < B < 4\times10^{12}$ for LRB, and $B > 2 \times 10^{13}$ for other models. In future, the long-term monitoring on these ULX pulsars in optical and X-ray bands would help to constrain the compact object masses.

At present, the limited number of ULX pulsars greatly restricts our understanding of their physical nature and accretion process. We used the currently available flux and period data, combined with different accretion torque models, to constrain the magnetic field and rotational behavior of the neutron stars. In addition, more precise period derivative data of know sources and new ULX pulsars are needed for further research. Present accretion torque models are based on the classical thin disc assumption, so considering the high-mass O- or early Be-type companions of ULX pulsars \cite{Kouroubatzakis2017,Bikmaev2017}, wind-fed models can also be possible options for ULX pulsars accretion form, and can be considered for future research.

\section*{Acknowledgements}
We are very grateful to the referee for the fruitful suggestions to improve the manuscript. This work is supported by the NSFC (U1838103, 11622326 and 11773008) and the National Program on Key Research and Development Project (Grants No. 2016YFA0400803).

\section*{Data Availability}
The data used in this paper were collected from the previous literatures, like refereed papers and the Astronomer's Telegrams. These data are public for all researchers.


\section*{References}
\bibliographystyle{elsarticle-num}
\bibliography{mybibfile}

\end{document}